\documentclass[aps,showpacs,floats,nofootinbib,
superscriptaddress]{revtex4}  
\usepackage{graphicx}

\newcommand{\glu}{{\tilde{g}}}

\newcommand{\vli}{\vec{\rm p}_{\ell_1}}
\newcommand{\vlii}{\vec{\rm p}_{\ell_2}}
\newcommand{\vbi}{\vec{\rm p}_{b_1}}
\newcommand{\vbii}{\vec{\rm p}_{b_2}}
\newcommand{\vnui}{\vec{\rm p}_{\tilde{\chi}^0_1}}
\newcommand{\pli}{p_{\ell_1}}
\newcommand{\plii}{p_{\ell_2}}
\newcommand{\pbi}{p_{b_1}}
\newcommand{\pbii}{p_{b_2}}
\newcommand{\pnui}{p_{\tilde{\chi}^0_1}}
\newcommand{\eli}{E_{\ell_1}}
\newcommand{\elii}{E_{\ell_2}}
\newcommand{\ebi}{E_{b_1}}
\newcommand{\ebii}{E_{b_2}}
\newcommand{\enui}{E_{\tilde{\chi}^0_1}}

\newcommand{\nuii}{\tilde{\chi}^0_2}
\newcommand{\nui}{\tilde{\chi}^0_1}
\newcommand{\mgl}{m_{\tilde{g}}}
\newcommand{\msb}{m_{\tilde{b}}}
\newcommand{\msbi}{m_{\tilde{b}_1}}
\newcommand{\msbii}{m_{\tilde{b}_2}}
\newcommand{\msl}{m_{\tilde{\ell}}}
\newcommand{\mnuii}{m_{\tilde{\chi}^0_2}}
\newcommand{\mnui}{m_{\tilde{\chi}^0_1}}
\newcommand{\xno}{{\rm x_1}}
\newcommand{\xni}{{\rm x_2}}
\newcommand{\xnii}{{\rm x_3}}
\begin{document}
\preprint{temp}
\title{A New SUSY mass reconstruction method at the CERN LHC}

\author{K. Kawagoe}
\affiliation{
Department of Physics, Kobe University, Kobe 657-8501, Japan
}
\author{M.M. Nojiri}
\affiliation{
YITP, Kyoto University, Kyoto 606-8502, Japan
}
\author{G. Polesello}
\affiliation{
CERN, CH-1211 Geneva 23, Switzerland and 
INFN, Sezione di Pavia, Via Bassi 6, 27100 Pavia, Italy
}
\date{\today}

\begin{abstract}
We propose a new mass reconstruction technique for SUSY processes at
the LHC.  The idea is to completely solve the kinematics of the SUSY
cascade decay by using the assumption that the selected events satisfy
the same mass shell conditions of the sparticles involved in the
cascade decay.  Using this technique, we study the measurement of the
mass of the bottom squarks in the cascade decay of the gluino.  Based
on the final state including two high $p_T$ leptons and two $b$-jets,
we investigate different possible approaches to the mass
reconstruction of the gluino and the two bottom squarks. In particular
we evaluate the performance of different algorithms in discriminating
two bottom squark states with a mass difference as low as 5\%.

\end{abstract}

\maketitle
\section{INTRODUCTION}

Supersymmetry is one of the most attractive models beyond the Standard
Model (SM) of elementary particle physics.  The superpartners of SM
particles (sparticles) might have masses of the order of the TeV, and
experiments at the Large Hadron Collider (LHC) should be able to
detect such particles up to masses of 2-3 TeV \cite{atltdr,
Abdullin:1998pm}.

The pattern of the sparticle mass spectrum depends on the
SUSY-breaking mechanism, which might depend on gravity, space-time
structure, or unknown interactions.  The unraveling of such mechanism
through the determination of the sparticles' masses is therefore one of
the most important physics targets of future collider experiments.

The potential of the LHC for SUSY mass determination has been studied
in detail in the past decade.  The most promising method involves the
study of the endpoints in the distributions 
of invariant masses among the visible sparticle decay products.  
Information on the masses involved in the cascade decay 
can be extracted from the
endpoint values if the decay distributions are dominated by a single
cascade decay chain.  Studies based on the endpoints are documented in
\cite{atltdr,Hinchliffe:1997iu,Hinchliffe:1999zc,Bachacou:1999zb,
Allanach:2000kt,Abdullin:1998pm,gjelsten}.

In this paper we explore a new method for reconstructing SUSY events.
This method does not rely only on events near the endpoint.  Instead,
one kinematically solves the neutralino momenta and masses of heavier
sparticles using measured jet and lepton momenta and optionally a few
mass inputs.

We concentrate in this paper on the measurement of the
mass of the bottom squarks (sbottoms) through the cascade decay: 
\begin{equation}
\tilde{g}\rightarrow \tilde{b}b_2\rightarrow \tilde{\chi}^0_2 b_1b_2
\rightarrow \tilde{\ell}b_1b_2\ell_2 \rightarrow \tilde{\chi}^0_1b_1b_2\ell_1\ell_2.
\label{bbll}
\end{equation}
We address this decay, rather than the equivalent cascade involving a
generic $\tilde q_L$ for various orders of reasons.  First of all, the
physics of the third generation squarks and leptons is particularly
important in disentangling the pattern of SUSY breaking. In the MSUGRA
model, $\tilde{b}_L$ is lighter than $\tilde{b}_R$ due to the RGE
running by top Yukawa coupling. The mass of the lighter sbottom state
$\tilde{b}_1$ is further reduced by the mixing of the left and right
$\tilde{b}$ states.  The sbottom sector is thus sensitive to
fundamental parameters of the theory such as the trilinear couplings
and $\tan\beta$ otherwise of difficult access at the LHC.  The third
generation sparticle masses are also important parameters for $B$ and
Higgs physics.  Second, the chain in Eq.~(\ref{bbll}) involves two $b$
quarks, which can be tagged in the detector.  The problem of correctly
identifying the jets contributing to the interesting decay is thus
made much simpler.  Finally, as we will discuss in detail below, both
sbottom states $\tilde b_1$ and $\tilde b_2$ yield the decay chain of
Eq.~(\ref{bbll}). The performance and the robustness of the
reconstruction algorithms can be benchmarked against the ability in
disentangling the two states.

Five sparticles are involved in the cascade decay Eq.~(\ref{bbll}),
therefore one can write five mass shell conditions among the leptons
and quarks in the final decay products.
\begin{eqnarray}
m^2_{\tilde{\chi}^0_1}&=& p^2_{\tilde{\chi}^0_1},\cr
m^2_{\tilde{\ell}}&=& (p_{\tilde{\chi}^0_1}+ p_{\ell_1})^2,\cr
m^2_{\tilde{\chi}^0_2}&=& (p_{\tilde{\chi}^0_1}+
p_{\ell_1}+p_{\ell_2})^2,\cr
m^2_{\tilde{b}}&=& (p_{\tilde{\chi}^0_1}+
p_{\ell_1}+p_{\ell_2}+p_{b_1})^2,\cr
m^2_{\tilde{g}}&=& (p_{\tilde{\chi}^0_1}+ 
p_{\ell_1}+p_{\ell_2}+p_{b_1}+p_{b_2})^2.
\label{gluino}
\end{eqnarray}

For a $bb\ell\ell$ event, the equations contain the 4 unknown degrees
of freedom of the $\tilde{\chi}^0_1$ momentum.  Each event therefore
describes a 4-dimensional hyper-surface in a 5-dimensional mass
parameter space, and the hyper-surface differs event by event. From
the purely mathematical point of view 5 events would be enough to
determine a discrete set of solutions for the masses of the involved
sparticles, and the probabilistic discussion of the following could be
easily developed in a 5-dimensional space. In order to illustrate the
method in a more transparent way, we will develop the argument by
assuming that the masses of $\tilde{\chi}^0_2$, $\tilde{\ell}$, and
$\tilde{\chi}^0_1$ are known.  This is a reasonable assumption at the
LHC, where it has been shown that a detailed study of the
lepton-lepton system from the $\tilde{\chi}^0_2$ decay can be used to
precisely constrain these masses \cite{Nojiri:2000wq}.  In this case,
each event corresponds to a different line in the $(\mgl,\msb)$ plane,
therefore two events are enough to solve the gluino and sbottom masses
altogether.

We call this technique the ``mass relation method'', because here one
uses the fact that sparticle masses are common for all events which go
though the same cascade decay chain.  Note that the events need not to
be close to the endpoint of the decay distribution, but they are still
relevant to the mass determination. This means that one can use the
mass relation method even if the number of signal events is small.

The purpose of this paper is to explore in detail the implications of
the mass relation method which is only sketched for signal events in
\cite{Nojiri:2003tu}.  In that report a measurement of the gluino and
sbottom masses was obtained from the peak of the distribution of the
solutions for all possible event pairs, assuming that the $\nuii$,
$\tilde{\ell}$ and $\nui$ masses are known.  In this paper we extend
the previous analysis to take SUSY backgrounds into account in the
distributions.

We note that $\tilde{b}$ in Eq.~(\ref{bbll}) could be either
$\tilde{b}_1$ or $\tilde{b}_2$.  The decay was studied in
\cite{gjelsten} by using the endpoint method, where the possibility of
distinguishing the two sbottom states $\tilde{b}_1$ and $\tilde{b}_2$
was studied for a case where the mass difference was approximately 5\%
of the sbottom mass. The conclusion showed that even with very large
integrated luminosity the result is at best marginal, and crucially
depends on the ability of the experimenters to model to a very high
level of detail the response function of the detector to $b$-jets.
This is probably an inescapable conclusion, given that the
resolution in calorimetric measurement is comparable to the mass
splitting one wants to measure. It is however worth studying alternative
reconstruction methods which make a better use of all the information
available for the selected events.

In a subsequent paper \cite{Lester}, a similar process is analyzed by
constructing an approximate event-likelihood function for signal
events, based on a Bayesian statistical approach.  In this paper we
construct a different approximation for the likelihood function. We
apply this function to the analysis of a simulated data set, and we
study how well $\tilde{b}_1$ and $\tilde{b}_2$ can be reconstructed
for a sample model point and its variants.

This paper is organized as follows. In section \ref{sec2}, we
introduce the detailed SUSY models addressed in our study, and briefly
present the simulation procedure. We then solve in section \ref{sec3}
analytically the kinematics for the process Eq.~(\ref{bbll}), arriving
to a compact expression for the masses of $\tilde{g}$ and $\tilde{b}$
as the function of $b$ and $\ell$ momentum and discuss the typical
solutions.  In section \ref{sec4}, we discuss the event pair analysis
on our sample SUSY model, where $m_{\tilde{g}}$ and $m_{\tilde{b}}$ are
computed from all event pairings  in the selected samples.  Gluino and
sbottom masses consistent with the input values are reconstructed, but
we are also forced to artificially select one of the multiple
solutions obtained from solving for the masses the coupled quadratic
equations for each event pair.  In section \ref{sec5}, we therefore
define an approximate likelihood function built using all the events
in the sample, and we describe an analysis based on this function.
The masses reconstructed from the likelihood analysis are in agreement
with the input values and the method automatically takes care of the
issue of multiple solutions without artificial selection.  We also
study the possibility to reconstruct the $\tilde{b}_2$ in our sample
model.  Section \ref{sec6} is devoted to discussions and comments.  We
especially discuss the theoretical relevance of being able to perform
detailed measurements in the third generation squark sector.

\section{Model points and simulations}
\label{sec2}
We choose for this study the model point SPS1a as defined in
Ref.~\cite{Allanach:2002nj} and its variants.  SPS1a has a significant
production cross-section for the chain of Eq.~(\ref{bbll}), and has
already been the subject of detailed LHC studies \cite{gjelsten}.  The
model is defined in the mSUGRA scenario by the parameters
$\tan\beta=10$, $m=100$~GeV, $M=250$~GeV, $A_0=-100$~GeV, and $\mu>0$.
In order to evaluate the performance of the analysis for different
values of the splitting of the two sbottom states, we also study the
points where $\tan\beta=15$ and $20$, keeping other GUT scale
parameters same as those for SPS1a.  For the additional points, the
$\glu$, $\nuii$, $\nui$ and $\tilde{\ell}$ masses are almost the same
as for SPS1a, while $\msbi$ is reduced because the left right mixing of
$\tilde{b}$ increases proportional to $\mu\tan\beta$.

The masses and decay branching ratios are calculated with the ISASUSY
code \cite{ISASUSY}.  The $m_{\tilde{b}_1}$ changes by 3\% from
$\tan\beta=10$ to $\tan\beta=20$.  We will see later that such 
a difference in the $\tilde{b}_1$ masses should be measurable 
at the LHC if systematic effects are kept under control.

Another drastic effect in Table~\ref{table2} is the decrease of
$Br(\tilde{\chi}^0_2\rightarrow \tilde{\ell}\ell )$ ($\ell=e,\mu$).
This is because the decay width for $\nuii\rightarrow \tilde\tau \tau$
increases as the $\tilde\tau$ mixing angle increases.  For
$\tan\beta=20$, the branching ratio into $bb\ell\ell$ mode is reduced
by a factor of 5 yielding a rather small number of accepted events.

We also list in Table~\ref{table2} the decay branching ratios relevant
to the decay chain of our interest.  The ratio of the decay branching
ratios $R_{\tilde{b}}=Br(\tilde{g}\rightarrow\tilde{b}_2
\rightarrow\tilde{\chi}^0_2)/
$$Br(\tilde{g}\rightarrow\tilde{b}_1\rightarrow\tilde{\chi}^0_2) $
varies from 0.26 to 0.14 when $\tan\beta$ changes from 10 to 20.  This
$\tan\beta$ dependence comes from the reduction of $\tilde{b}_1$ mass
and the increase of the left-right mixing of $\tilde{b}$, $\theta$,
defined as $\tilde{b}_1\equiv \cos\theta
\tilde{b}_L+\sin\theta\tilde{b}_R$, from $\theta=0.49$
($\tan\beta=10$) to 0.61 ($\tan\beta=20$).  Namely,
\begin{itemize}
\item $\Gamma(\tilde{g}\rightarrow\tilde{b}_1) $ increases 
as $m_{\tilde{b}_1}$ is reduced. 
\item   $\Gamma(\tilde{b}_2\rightarrow\tilde{\chi}^0_2)$ 
increases as $\theta$ increases. 
However, $\Gamma(\tilde{b}_2\rightarrow\tilde{\chi}^+_1)$ and 
$\Gamma(\tilde{b}_2\rightarrow\tilde{t}_1)$ also increase, 
therefore
$Br (\tilde{b}_2\rightarrow\tilde{\chi}^0_2)$ stays more or less same. 
\item $m_{\tilde{b}_1}$ is reduced and 
$\Gamma(\tilde{b}_1\rightarrow \tilde{t}_1W)$ 
is kinematically suppressed as $\tan\beta$ increases.
\end{itemize}
The branching ratio is in principle important information 
to determine the mixing angle $\theta$ but one needs to 
know the relevant sparticle mass spectrum to utilize 
it. 

Note that the ratio $R_{\tilde{b}}$ is small and the mass splitting
between $\tilde{b}_1$ and $\tilde{b}_2$ is within 10\% over the
parameters we study.  The $\tilde{b}_1$ is therefore significant
background for the $\tilde{b}_2$ search at the LHC.  Nevertheless, a
hint for the decay of $\tilde{g}\rightarrow
\tilde{b}_2$ might still be observable in the $bb\ell\ell$ distribution, 
given an excellent control of the detector response to $b$-jets, 
as we will see in section~\ref{sec5}.

The SUSY events are generated using the HERWIG 6.4
\cite{Corcella:2000bw,Moretti:2002eu} event generator.  The produced
events are passed through the ATLFAST \cite{ATLFAST} detector
simulator, which parameterizes the response of the ATLAS detector. In
particular, we use the parametrization for $b$-tagging efficiency
corresponding to the expected high luminosity performance of the ATLAS
detector.  While the performance is a function of the $p_T$ of the
jets, a typical performance figure is $\epsilon_{b}=0.5$ for the $b$
tagging efficiency for a rejection of 100 on light quark jets.

For each point we have generated $1.5\times 10^7$ events which 
approximately correspond to an integrated luminosity of 
300~{fb}$^{-1}$.

The following cuts are applied in order 
to select signal events:
\begin{itemize}
\item $M_{\rm eff}>600$~GeV
and $E_T^{\rm miss}> 0.2 M_{\rm eff}$, where $E_T^{\rm miss}$ is the
missing transverse energy and $M_{\rm eff}$ is the scalar sum of the missing
transverse energy and the transverse momenta of the four hardest jets,
\item at least 3 jets with $p_{T1}>150$~GeV, $p_{T2}>100$~GeV and
$p_{T3}>50$~GeV,
\item exactly two $b$-tagged jets with $p_T>50$~GeV,
\item exactly two opposite-sign isolated same-flavor
(OSSF) leptons with $p_{T\ell_1}>20$~GeV, 
$p_{T\ell_2}>10$~GeV,
with an 
invariant mass $40$~GeV$<m_{\ell\ell}<78$~GeV.
\end{itemize}

The isolation criterion consists in requiring a transverse energy
deposition in the calorimeters smaller than $10$~GeV in a
($\eta,\phi$) cone of radius $0.2$ around the lepton direction, where
$\eta$ is the pseudorapidity of the lepton and $\phi$ the angle in the
plane transverse to the beam.  A detailed discussion of the Standard
Model backgrounds after these cuts is given in \cite{gjelsten}. The
authors show that the Standard Model background is negligible in
comparison to the SUSY background, consisting of SUSY events not
including the decay chain of Eq.~(\ref{bbll}).

In order to perform the reconstruction, we need to identify the
position of each of the two $b$-jets and of each of the two leptons in
the decay chain.  In the following analysis, we assume that the
$b$-jet with larger $p_T$ originates from the $\tilde{b}$ decay. The
assignment is optimal for SPS1a, because $\msb-\mnuii\gg\mgl-\msb$.
At the same time one can fix the lepton assignment so that the higher
(lower) $p_T$ lepton $\ell_{\rm high}$ ($\ell_{\rm low}$) comes from
$\tilde{\ell}$ to increase the possibility to pick up the correct
lepton assignment.  If we roughly know the masses of $\tilde{\ell},
\nui$ and $\nuii$, it is easy to determine which assignment is
optimal~\cite{Nojiri:2000wq}. For the likelihood analysis on SPS1a, we
use $\ell_{\rm high}$ as $\ell_1$ in Eq.~(\ref{bbll}).

\begin{table}
\begin{center}
\begin{tabular}{|c|c|c|c|c|c|}
\hline
 $\tan\beta$ &
$m_{\tilde{g}}$ & $m_{\tilde{b}_{1(2)}}$ & $m_{\tilde{\chi}^0_2}$ & 
$m_{\tilde{\ell}_R}$ 
&$m_{\tilde{\chi}^0_1}$ \cr
\hline
10 & 595.2& 491.9 (524.6)& 176.8&  143.0& 96.0\cr
\hline
15 & 595.2& 485.3 (526.9)& 177.9&  143.0& 96.5\cr
\hline
20 & 595.2& 478.7 (531.2)& 178.5&  143.1& 96.7\cr
\hline
\end{tabular}
\caption{Masses of the relevant sparticles for the three studied 
points, in GeV.}
\end{center}
\label{table1}
\end{table}

\begin{table}
\begin{center}
\begin{tabular}{|c|c|c|c|}
\hline
 $\tan\beta$ &
$Br(\glu \to \tilde{b}_{1(2)})$ &
$Br(\tilde{b}_{1(2)} \to \tilde{\chi}_2^0)$ &
$Br(\tilde{\chi}_2^0 \to \tilde{\ell})$\cr
\hline
 10 &
8.9 (4.9) $\times$2 &
35.8 (16.7) &
3.16$\times$4\cr
\hline
 15 & 
9.8 (4.6) $\times$2 &
37.8 (15.9) &
1.26$\times$4\cr
\hline
 20 &
10.8 (4.1) $\times$2 &
38.9 (13.2) &
0.61$\times$4\cr
\hline
\end{tabular}
\caption{Branching fractions for the decays used in the 
analysis, in percent.}
\end{center}
\label{table2}
\end{table}

\section{Solutions of the decay kinematics}
\label{sec3}

\subsection{Formula of the decay kinematics}
It is straightforward to solve the decay process in Eq.~(\ref{bbll}).
We first note that the $\tilde{b}$ cascade decay can be written as a
function of momenta of $b_1$, $\ell_1$ and $\ell_2$ , because the four
mass shell conditions can be used to eliminate the unknown $\nui$ four
momentum.

To systematically solve the system, we expand the $\nui$ momentum 
with the observed momenta of $b_1$, $\ell_1$, and $\ell_2$:
\begin{equation}
\vnui=a \vli+ b \vlii+ c\vbi.
\label{expand}
\end{equation}
The expansion is possible 
if $\vli$, $\vlii$ and $\vbi$ are independent from one another. 
The three on-shell conditions may be then rewritten as 
\begin{equation}
{\cal M}\left(\begin{array}{c}a\cr b\cr c\end{array}\right)
=\left(\begin{array}{lll}
-\frac{1}{2}(\msl^2-\mnui^2)&&+ \eli\enui \cr
-\frac{1}{2}(\mnuii^2-\msl^2-2\pli\cdot\plii)&&+\elii\enui \cr
\frac{1}{2}\left(\mnuii^2+m_b^2+2\pbi\cdot(\pli+\plii)\right)& 
-\frac{1}{2}\msb^2& +\ebi\enui\cr
\end{array}
\right)
\end{equation}
where 
\begin{equation}
{\cal M}= \left(
\begin{array}{ccc}
\vli\cdot\vli &\vli\cdot\vlii& \vli\cdot\vbi\cr
\vli\cdot\vlii & \vlii\cdot\vlii & \vlii \cdot\vbi\cr
\vli \cdot \vbi & \vlii\cdot \vbi & \vbi\cdot\vbi\cr 
\end{array}
\right).
\end{equation}

The three parameters, $a,b$ and $c$ are solved as functions 
of $\enui$ if $\det {\cal M}\neq 0$. By using the on-shell 
condition of $\tilde{\chi}^0_1$, we obtain the quadratic 
equation of $\enui$ as 
\begin{equation}
A_{33}\left(\frac{\enui}{\mnui}\right)^2+ 
2\left(A_{13} + \frac{\msb^2}{\mnui^2} A_{23}\right)
\left(\frac{\enui}{\mnui}\right)
+ \left(A_{11}+ 2 \frac{\msb^2}{\mnui^2} A_{12}+ 
\frac{\msb^4}{\mnui^4}A_{22}\right)
=0   
\end{equation}
where $A_{ij} = [{\rm x_i}]^T {\cal M}^{-1} [{\rm
x_j}]-\delta_{ij}(i-2)/\mnui^2$ and definition of ${\rm x_1}$
is given in the appendix.

As the $\tilde{b}$ decay kinematics is solved,
we now use the on-shell condition of $\tilde{g}\rightarrow \tilde{b}b_2$ 
to obtain 
\begin{equation}
Q_{11} \mgl^4+2Q_{12} \mgl^2 \msb^2+ Q_{22}\msb^4
+2Q_{1} \mgl^2+ 2Q_{2}\msb^2 + Q=0.
\label{glsb}
\end{equation} 
$Q$'s are  functions of the momenta of the leptons 
and $b$ quarks, $\mnuii$, $\mnui$, and $\msl$. 
The expressions for $Q$'s are shown in Appendix.

Mathematically, when there are two independent events, we have two
independent equations of the form Eq.~(\ref{glsb}).  The equation is
quadratic for $\mgl^2$ and $\msb^2$ and can be analytically
solved\footnote{Note that when some of the momenta are parallel,
${\cal M}$ cannot be inverted. We ignore the possibility, because
experimentally we always require the isolation of leptons from jets.}.
There are up to four solutions for an event pair, and one of them must
coincide with the true solution.

\section{Endpoint analysis and mass relation method}

The decay chain $ 
\tilde{g}\rightarrow \tilde{b}b_2\rightarrow \tilde{\chi}^0_2 b_1b_2
\rightarrow \tilde{\ell}b_1b_2\ell_2 \rightarrow 
\tilde{\chi}^0_1b_1b_2\ell_1\ell_2
$ has already been studied in detail in Refs.~\cite{gjelsten,
massimiliano}, using the endpoint method shown in Ref.~\cite{atltdr}.
In this method, the masses of the $\tilde{\chi}^0_1$ and
$\tilde{\chi}^0_2$ are assumed to be known from the study of the
kinematic edges in the $\tilde{q}_L$ decay. For events near the edge
of the $m_{\ell\ell}$ distribution the $\tilde{\chi}^0_1$ is
essentially at rest, and the momentum of the $\tilde{\chi}^0_2$ can be
approximated with the relation:
\begin{equation}
{\bf p}_{\tilde{\chi}^0_2} \simeq
\left(1-\frac{m_{\tilde{\chi}^0_1}}{m_{\ell\ell}}\right){\bf p}_{\ell\ell}.
\label{wrong}
\end{equation}

\begin{figure}[thb]
\begin{center}
\includegraphics[width=6.5cm]{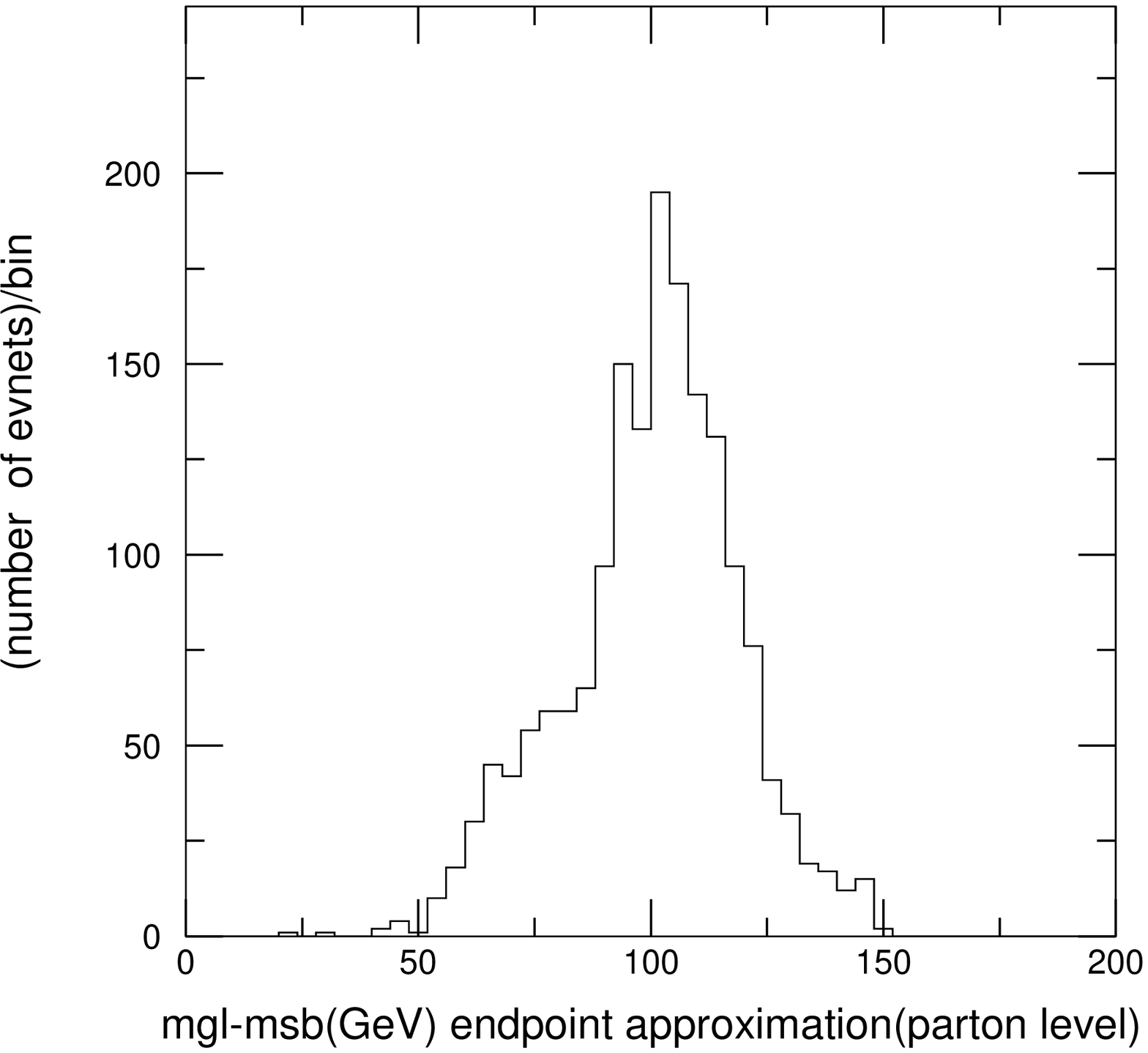}
\includegraphics[width=6.5cm]{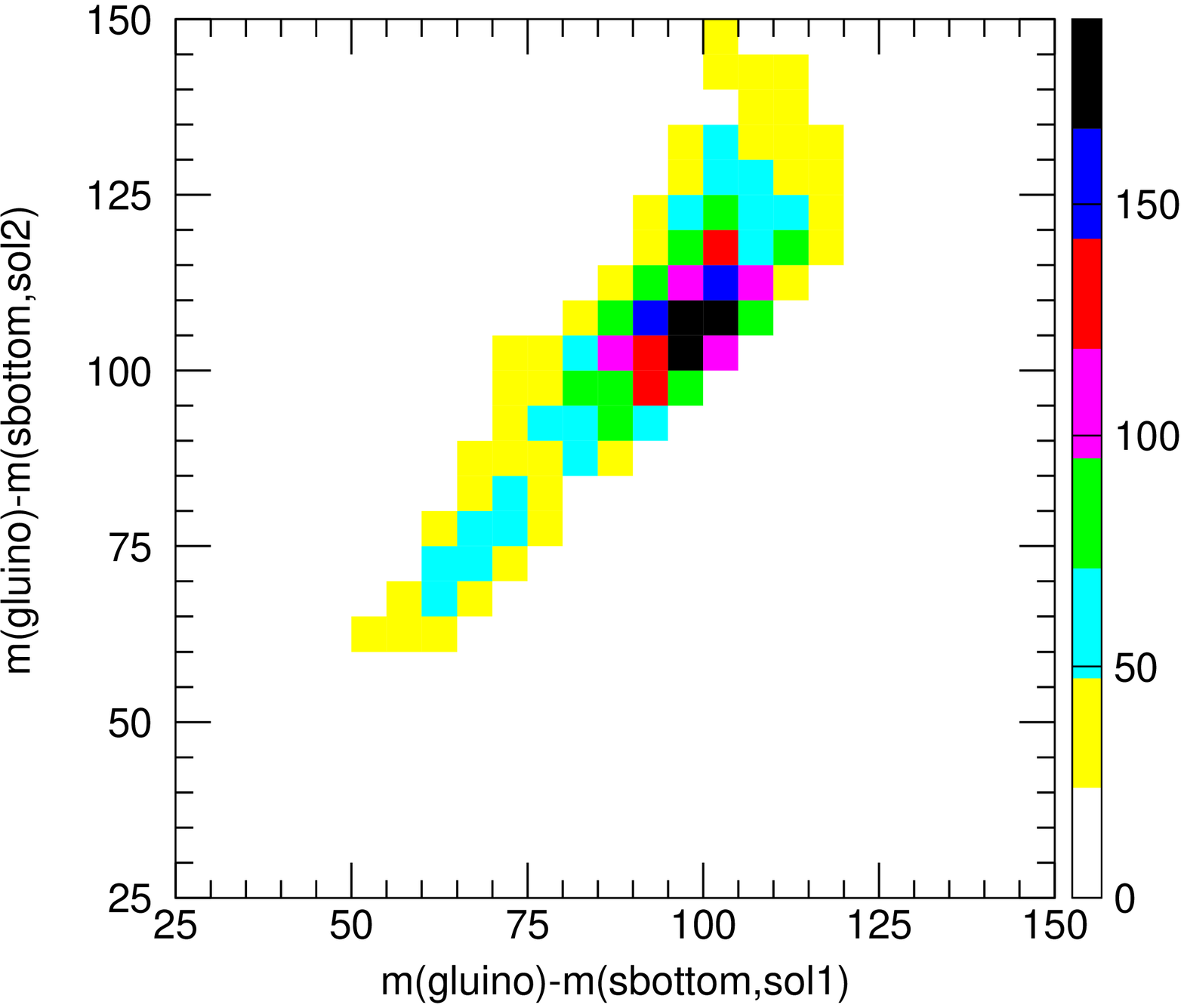}
\end{center}
\begin{center}
a)\hskip 5cm b)
\end{center}
\caption{The distribution of $m_{\tilde{g}}- m_{\tilde{b}}$ calculated
using a) parton level $b$ momentum and using the approximate 
relation Eq.~(\ref{wrong}) and  b) by solving Eq.~(\ref{gluino}). 
For a) $m_{\ell\ell}>65$ GeV is required. 
For b), the two $m_{\tilde{b}}$ solutions for the input gluino mass 
of $\mgl=595$~GeV 
are plotted in a two-dimensional plane, for 
both  correct and wrong lepton assignment. }
\label{parton}
\end{figure}

This formula is correct only at the endpoint of the three body decay
$\tilde{\chi}^0_2\rightarrow\chi^0_1 \ell\ell$, but is nevertheless
approximately correct near the edge of $\tilde{\chi}^0_2\rightarrow
\tilde{\ell}\ell \rightarrow \ell\ell\tilde{\chi}^0_1$ at SPS1a.  The
sbottom mass can then be calculated by building the invariant mass of
the approximate $\tilde{\chi}^0_2$ obtained by using Eq.~(\ref{wrong})
with the leading $b$-jet in the event.  The parton-level result is
shown in Fig.~\ref{parton}(a), where we plot the difference between
the reconstructed gluino mass and the reconstructed sbottom mass, to
minimize the smearing introduced by the approximation of
Eq.~(\ref{wrong}).  The $\tilde{b}_1$ peak is reconstructed correctly.
The bump on the left originates from the events from
$\tilde{g}\rightarrow\tilde{b}_2$, and is centered on the position
$\mgl-\msbii=$70.6~GeV. Even at parton level, with no experimental
smearing the bump is not well separated from the $\tilde{b}_1$ peak at
103~GeV.

Unlike the relation Eq.~(\ref{wrong}), the formula Eq.~(\ref{glsb}) is exact. 
For each event we have in this case two possible lepton assignments,
and for each lepton assignment, given an input value for the gluino 
mass, two solutions for the sbottom mass from the quadratic 
equation Eq.~(\ref{glsb}).
We show in Fig.~\ref{parton} the smaller solution  versus the larger one
of Eq.~(\ref{glsb}) for both of the lepton assignments. For the 
gluino mass the nominal value is assumed.
Unlike in Fig.~\ref{parton} a) two peaks of 
$\tilde{b}_1$ and $\tilde{b}_2$ are clearly separated in the plot.
Furthermore the number of the available events for the  mass fit 
are now increased by factor of 2 compared to the endpoint 
analysis because there are no constraint on the value of 
$m_{\ell\ell}$  when using Eq.~(\ref{glsb}). 
\par
The advantage of switching to the exact solution
for the event kinematics is clearly demonstrated by 
the plots. In order to evaluate if the heavier sbottom state
will be detectable we need to perform the study taking into account
the smearing induced by the fact that $b$-jets rather than partons are
measured in the detector.

\section{Event pair analysis}
\label{sec4}

As discussed in section \ref{sec3}, each event can be represented as a
quadratic equation in the $\mgl^2$, $\msb^2$ variables.  By taking two
events, we have a system of two equations in two unknowns which can be
solved analytically. This yields up to four values for the squark and
gluino masses. In the following, We start from the selected
$bb\ell\ell$ events, and we build all the possible event pairs.  In
order to minimize the combinatorial backgrounds we use the pairings
which satisfy the following conditions
\footnote{Note the selections  are rather 
phenomenological and they may introduce some bias to the reconstructed 
sparticle masses. We find however the obtained peak positions are
consistent with the 
input masses in this study.}:
\begin{itemize}
\item Eq.~(\ref{glsb})
has solution for only one of the two possible lepton assignments.
\item
For the selected lepton assignment the resulting quartic equation in
$\mgl^2$ has only two solutions, and the difference of the gluino
masses for the two solutions is more than 100~GeV.  The smaller gluino
mass solution is chosen.
\end{itemize}

\begin{figure}[thb]
\centerline{
\includegraphics[width=6cm]{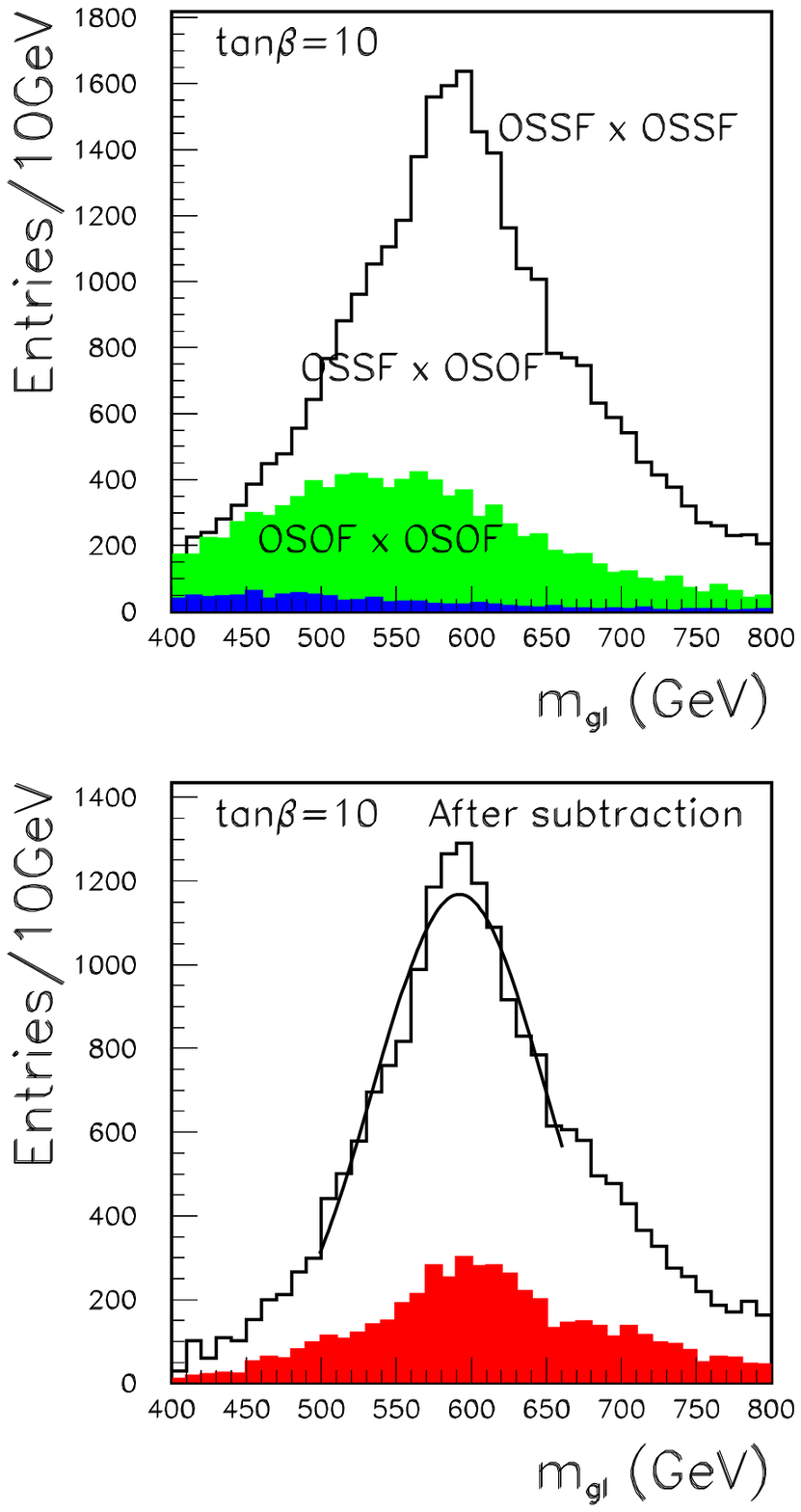}
\includegraphics[width=6cm]{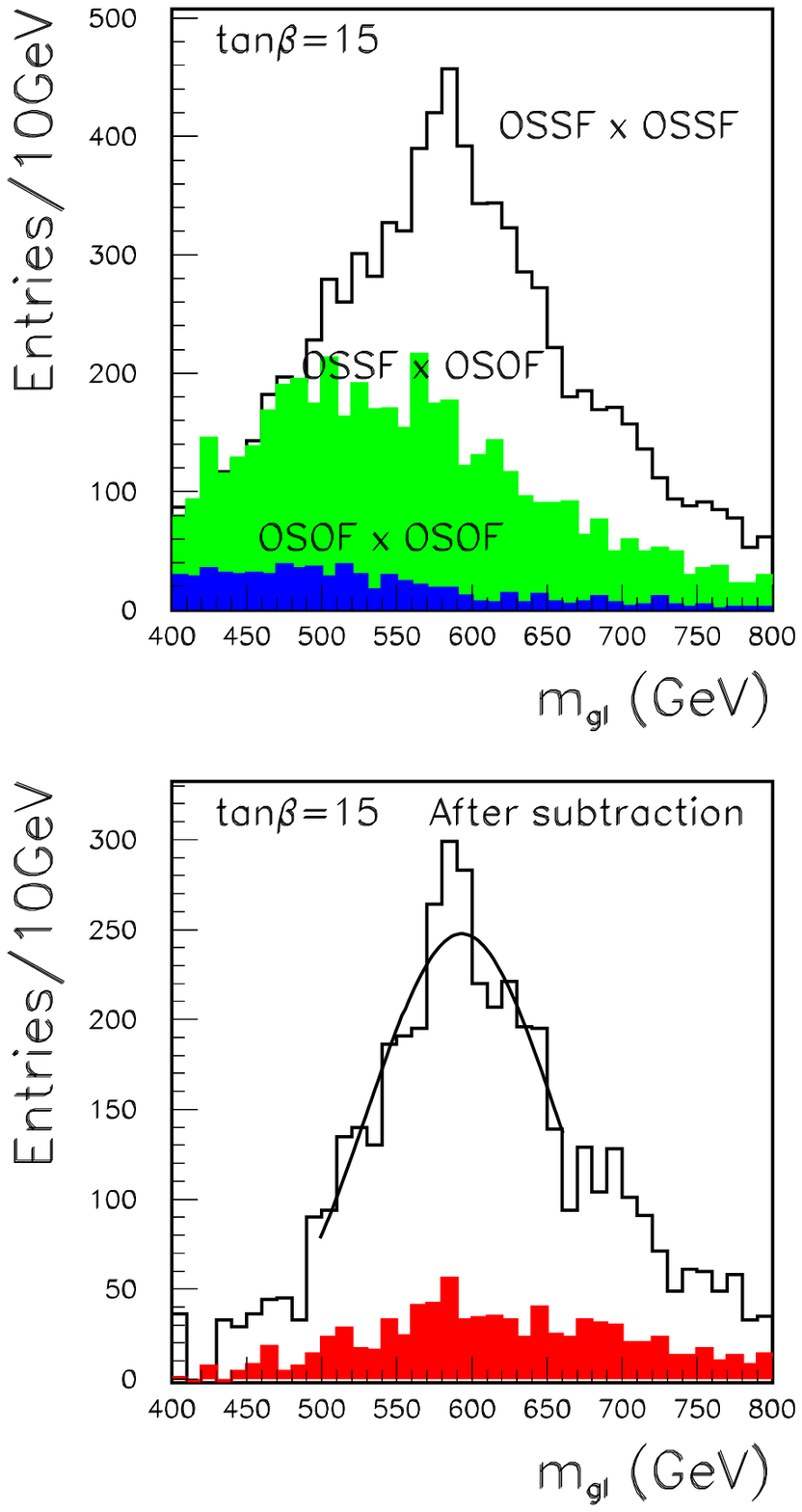}
\includegraphics[width=6cm]{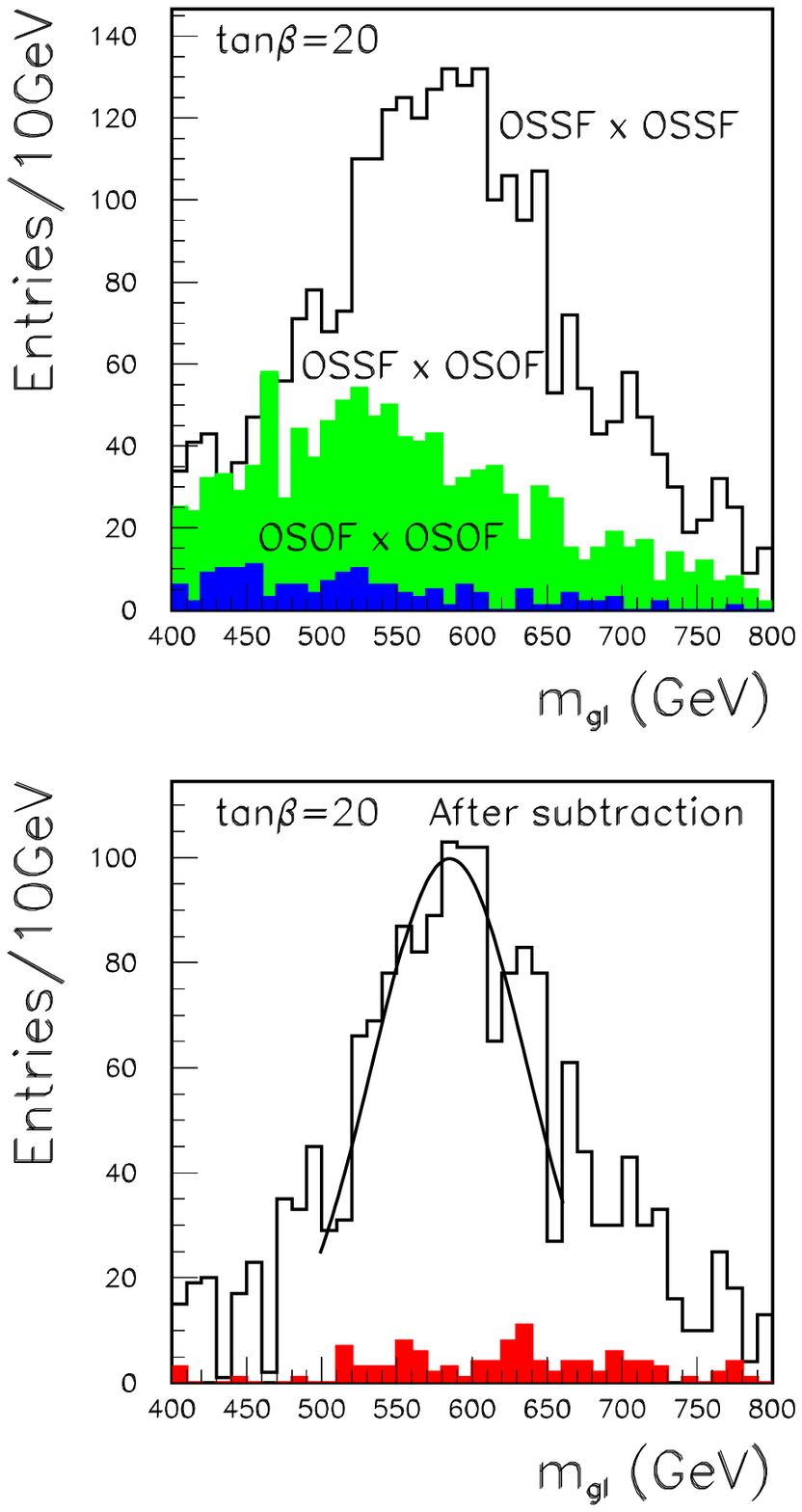}
}
\caption{The gluino mass distributions for
$\tan\beta=10$ (left),
$\tan\beta=15$ (center)
and $\tan\beta=20$ (right) with the event pair analysis.
The open, green, and blue histograms
in the top figures are for OSSF$\times$OSSF,
OSSF$\times$OSOF, and OSOF$\times$OSOF event pairs,
respectively.
The open histograms in the bottom figures
shows the mass distributions after background subtraction.
The contributions of $\tilde{b}_2$ are shown by red histograms.
}
\label{mglplot}
\end{figure}

The $\mgl$ distributions for the OSSF$\times$OSSF events pairs are
shown in the histograms on the upper line of Fig.~\ref{mglplot}.  A
significant SUSY background, also shown in Fig.~\ref{mglplot} is still
present in the sample.  This background can be estimated from the data
themselves by using the $bb\ell\ell$ events with an opposite sign
opposite flavor (OSOF) lepton pair (i.e. $\ell\ell=e^{\pm}\mu^{\mp}$).
To this purpose we produce mass distributions for the three types of
event pairs:
\begin{enumerate}
\item
two OSSF lepton events (OSSF$\times$OSSF),
\item
an OSSF lepton event and an OSOF lepton event (OSSF$\times$OSOF),
\item
two OSOF lepton events (OSOF$\times$OSOF).
\end{enumerate}
The background-subtracted distribution
can be then obtained as the combination of the three distributions:
OSSF$\times$OSSF$-$OSSF$\times$OSOF$+$OSOF$\times$OSOF.

The total number of event pairs is of course much larger than the
number of events.  Due to the selection criteria imposed to minimize
the combinatorial backgrounds, some of the events are not used at all
to make event pairs, while events which are used at least once are
used O(10) times on average.  The three histograms show peaks
corresponding to the input value for the gluino mass even before the
background subtraction.  The green and blue histograms show the
OSSF$\times$OSOF and OSOF$\times$OSOF distributions, respectively.
The distributions after the background subtraction are shown in the
histograms on the lower line of Fig.~\ref{mglplot}.  The peak position
and its error obtained by a Gaussian fit to the distribution are
listed in Table~\ref{mglfit}.  We also show in the histogram the
gluino mass distribution for event pairs where at least one of the
events comes from $\tilde{b}_2$ decay. The $\tilde{b}_2$ contribution
is significantly smaller compared with the distribution of
$\tilde{b}_1$ pair and does not affect the gluino mass fit.
One can also look into the distribution of
$m_{\tilde{g}}-m_{\tilde{b}}$ and estimating the value of this
observable by performing a Gaussian fit on the observed peak.  The
result of the fit is also shown in Table~\ref{mglfit}.

The statistical error of the gluino mass measurement can be evaluated
by performing the analysis on a set of statistically independent 
experiments.
To perform the evaluation within a reasonable
CPU budget, we generated a set of events where the $\tilde{b}_1$
is forced to decay with 100\% BR into the desired decay chain, for the 
SPS1a Point ($\tan\beta=10$).
The generated statistics corresponds to 30 experiments with an
integrated luminosity of 300~fb$^{-1}$ each.
By performing the analysis on the 30 experiments,
we find that the spread for the measured gluino mass is 1.6~GeV for
the benchmark integrated luminosity of 300~fb$^{-1}$.
By construction, the presence of the combinatorial background is not 
considered in the error analysis. No large degradation of the
resolution is expected once this effect is correctly taken
into account.

\begin{figure}[thb]
\centerline{
\includegraphics[width=6cm]{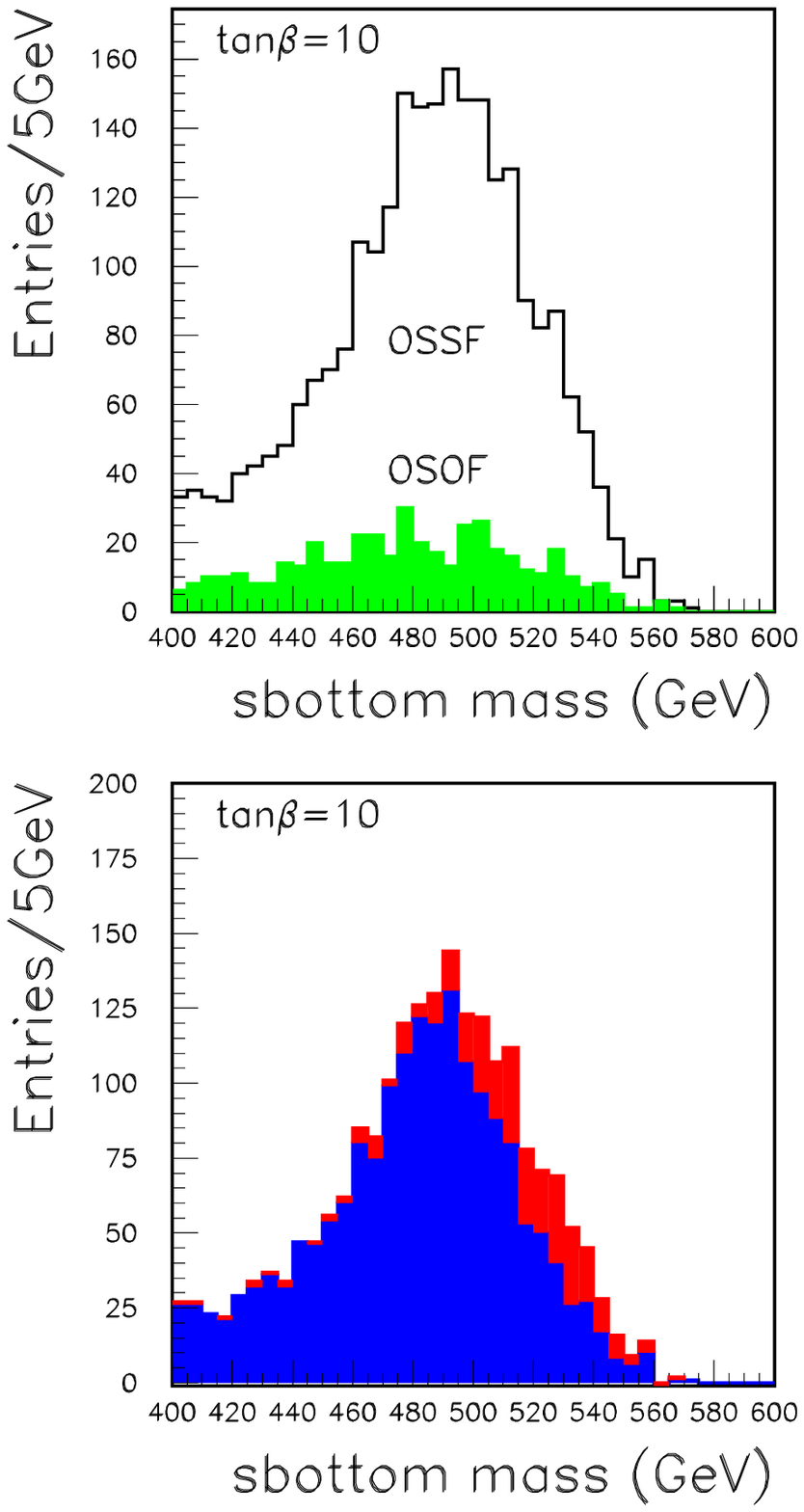}
\includegraphics[width=6cm]{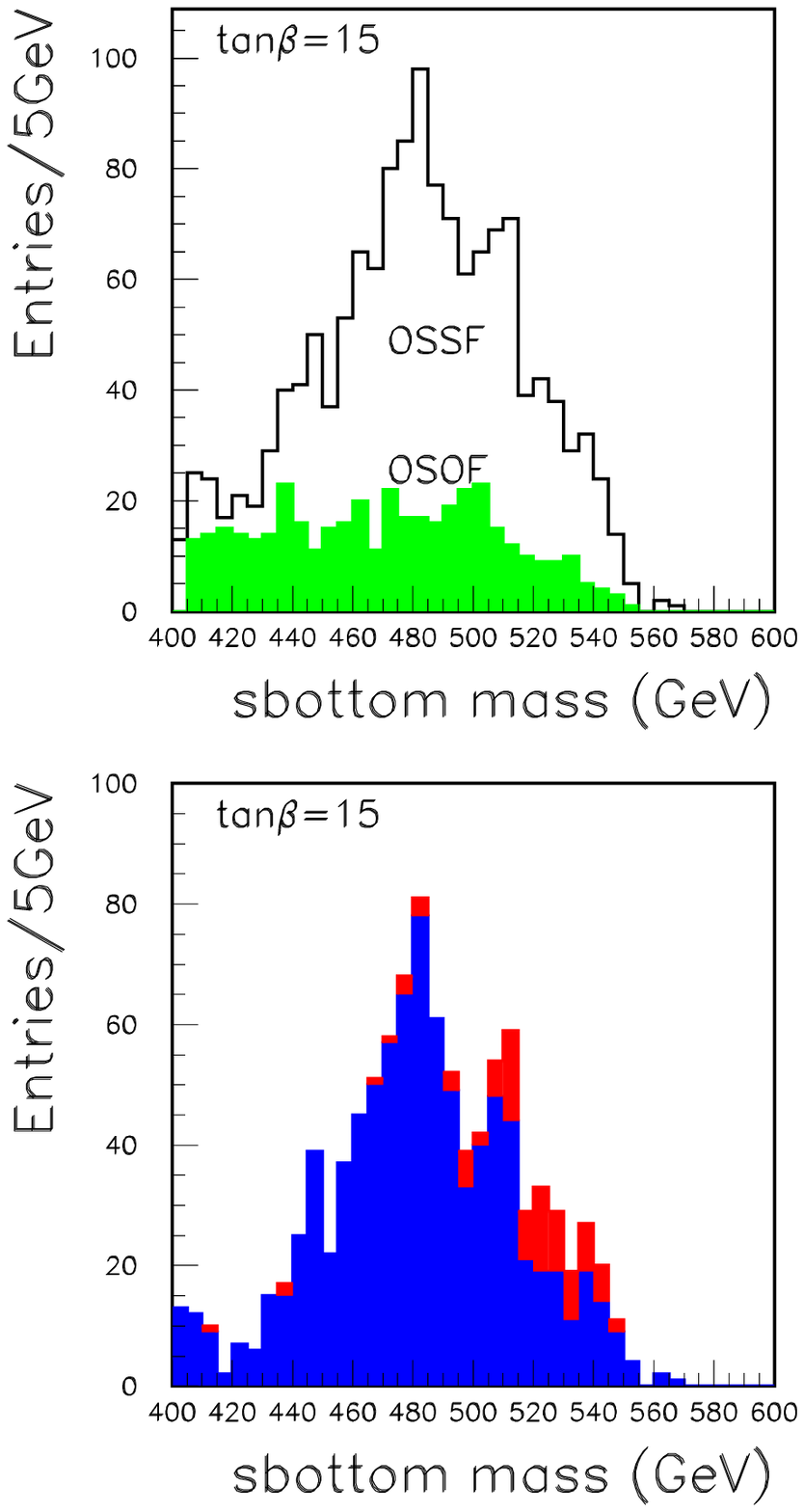}
\includegraphics[width=6cm]{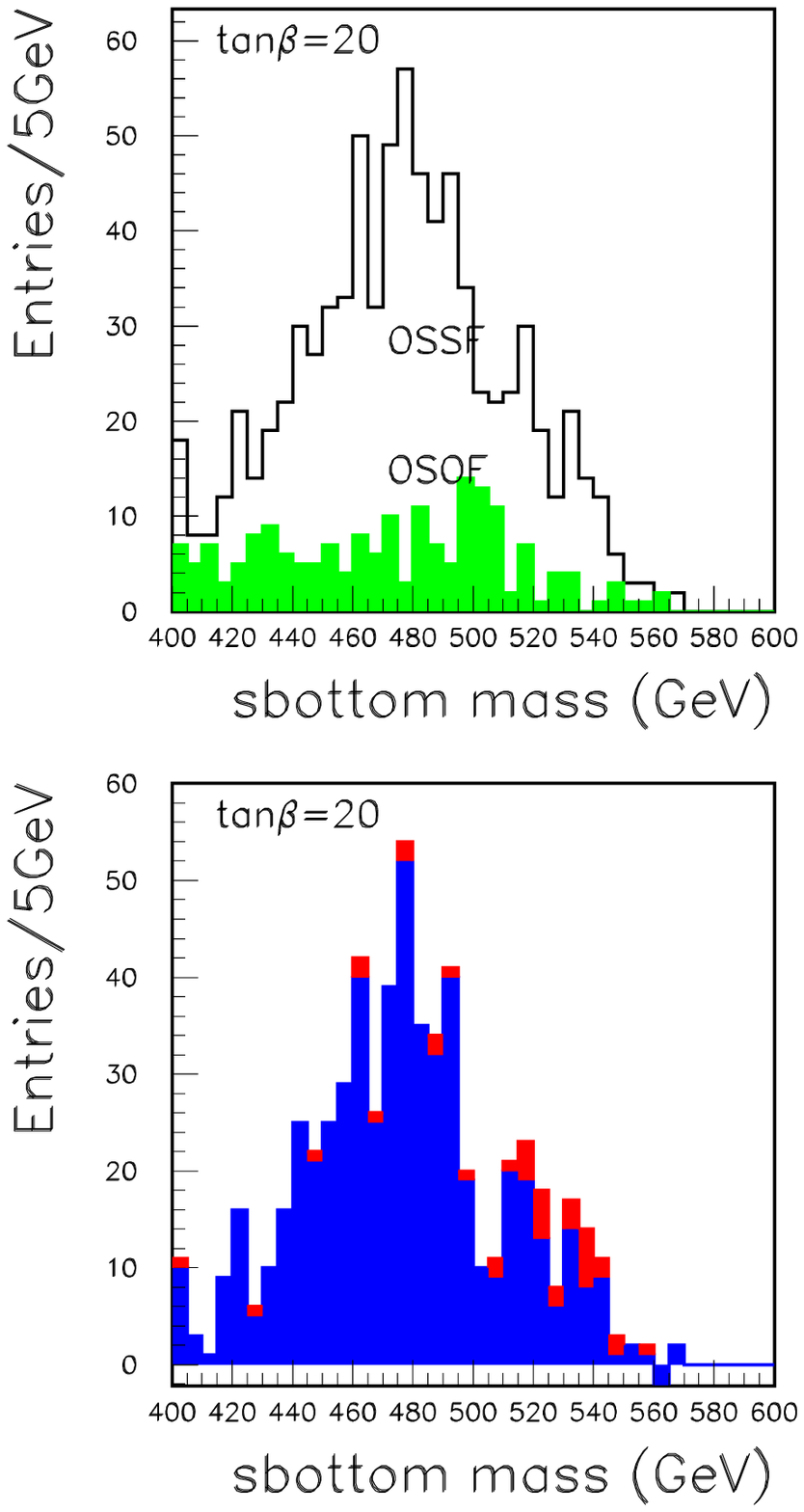}
}
\caption{The $m_{\tilde{b}}$ distributions
for $\tan\beta=10$ (left), 15 (center)  and 20 (right)
with a fixed gluino mass
($\mgl=595$~GeV).
The open and green histograms in the top figures
show the distributions of OSSF and OSOF lepton events, respectively.
The mass distributions after background subtraction
are shown in the bottom figures,
where the contribution of the $\tilde{b}_2$ events are shown
by the red regions.
}
\label{smallmsb}
\end{figure}

Once the gluino mass is fixed by the analysis shown above,
Eq.~(\ref{glsb}) can be solved for each event for the sbottom mass,
giving as input the central measured value for the gluino mass.
Following this procedure, in Fig.~ \ref{smallmsb}, we plot the
distribution of the smaller sbottom mass solution $m_{\tilde{b}}$(min)
for both OSSF (signal) and OSOF (background) lepton pair events (top
histograms).  We show in the bottom line the mass distributions after
the background subtraction, where the red regions show the
contribution of $\tilde{b}_2$.  The peak positions are evaluated by a
Gaussian fit, and listed in Table~\ref{mglfit2}.  Note that the total
number of the signal $\tilde{b}_1$ event is smaller by factor of 4 for
$\tan\beta=20$ compared with $\tan\beta=10$, but the mass peak is seen
very clearly.  The $m_{\tilde{b}}$(min) peak and
$m_{\tilde{b}}$(input) are in good agreement, and it is not so for the
larger solution $m_{\tilde{b}}$(max).

The peak positions of the distribution of the events originated from
$\tilde{b}_2$ decay are also consistent with $\tilde{b}_2$ masses. It
is however evident from the plots that in the real experiment it will
not be possible to claim the presence of the second peak. The
existence of $\tilde{b}_2$ can be established only after understanding
$b$-jet smearing and $\tilde{b}_1$ distribution correctly.

\begin{table}
\begin{tabular}{|c||c|c|c|}
\hline
            &$\tan\beta=10$&   $\tan\beta=15$&   $\tan\beta=20$\cr
\hline
$\mgl$& 591.9&593.1& 585.1\cr
$\mgl-\msb$ & 98.9 &  105.1 & 111.6 \cr
\hline
\end{tabular}
\caption{Fit results of the gluino and and sbottom masses in
the event pair analysis.
}
\label{mglfit}
\end{table}

\begin{table}
\begin{tabular}{|c||c|c|c|}
\hline
            &$\tan\beta=10$&   $\tan\beta=15$&   $\tan\beta=20$\cr
$\msb$(true)    &      491    &          485.3  &          478.8\cr
$\msb$(min)&  492.1$\pm$ 1.2& 487.7$\pm$ 2.2& 474.3$\pm$ 2.4\cr
$\msb$(max)&  504.5$\pm$ 1.0& 502.9$\pm$ 1.7& 495.1$\pm$ 2.4\cr
\hline
\end{tabular}
\caption{Fit results of the sbottom mass with a  fixed gluino mass
($\mgl=595$~GeV).
}
\label{mglfit2}
\end{table}

\section{Likelihood analysis}
\label{sec5}

The analysis shown in the previous section only uses
the events in pairs to evaluate the values of the squark
and gluino masses. A more efficient use of the available
event statistics can be achieved by using all of the
events at the same time and finding the ($\mgl, \msb$) pair 
for which the combined probability of all events is highest.
We define in this section an approximate likelihood function
for  ($\mgl, \msb$), and we then apply it to detection
of $\tilde{b}_2$.

\subsection{Construction of the likelihood function} From 
Eq.~(\ref{glsb}), each event is represented as 
a curve in the ($\mgl, \msb$) plane. The coefficients of the curve are 
a function of  the four momenta of the detected partons. 
The partons are measured as jets in the detector, which smears
the parton according to a smearing function.  It depends
on the detector performance and on the algorithms used
to cluster the energy deposition in the detector.\par
Due to these experimental effects, in a frequentist approach,
from the measured quadratic form for an event we can build a 
confidence belt in the ($\mgl, \msb$) plane.
This should be built using a Neyman construction \cite{Hagiwara:2002fs}, 
from the probability distribution for the measured values of 
($\mgl, \msb$) as a function of the input values for the
same two variables.
In order to build this function, the crucial ingredient is 
the distribution of the measured $b$-jet momenta as a function 
of the $b$-parton momenta. Evaluating this distribution
is outside
the scope of this work, as it requires a detailed simulation of
the detector response, which will need to be validated on 
real data using calibration samples of $b$-jets in the detector.
Even assuming an approximate form for the $b$-jet response, 
the proper Neyman construction for each event is a very computing-intensive
calculation. \par
We recast equation Eq.~(\ref{glsb})  in the form
$$
f(\mgl,\msb,p_1,p_2)=0
$$
where $p_1$ and $p_2$ are the momenta of the two $b$-jets.
We build an 
approximate probability density function according to the formula:
\begin{equation}
{\cal L}(\mgl,\msb)
=\int dp'_1 \int dp'_2 \epsilon(p_1:p'_1)\epsilon(p_2:p'_2)
\delta(f(\mgl,\msb,p'_1,p'_2))
\label{like}
\end{equation}
where $\epsilon(p:p')$ is the probability to measure a momentum $p'$ 
for a $b$-jet, given a $b$-parton with momentum $p$. 
%The function $F(\mgl,\msb,p'_1,p'_2)$ takes the value 1 when 
%$f(\mgl,\msb,p_1,p_2)=0$ and the value 0 otherwise.

In the equation we did not include the possibility of 
lepton momentum mis-measurement, which has an almost negligible
effect, as compared to the smearing of $b$-partons.  
As a further simplification,
we assume that the jet direction is not 
modified by the measurement and we use for $\epsilon(p:p')$ a gaussian 
distribution, with a width $\sigma$ 
corresponding to the parameterized jet smearing 
used in the fast simulation program. 
\begin{eqnarray}
\sigma/E&=&0.5/\sqrt{E{\rm (GeV)}}+0.03\ \ \  (\vert\eta\vert<3.0)\cr
\sigma/E&=&1.0/\sqrt{E{\rm (GeV)}}+0.07\ \ \  (\vert\eta\vert>3.0)
\end{eqnarray}
The gaussian smearing  is not very good approximation for 
for $b$-jets for which in many cases the semi-leptonic decay
of the $b$-quark results in jets containing an unmeasured
neutrino.  
The approximate function takes however into account the
dominant part of the jet smearing and can be used to demonstrate the
method. Note also that  in 
Eq.~\ref{like}, we define our ${\cal L}$ using
$\epsilon(p_1,p_1')$  where 
$p_1$ is measured $b$-jet momentum. 
The function ${\cal L}$ would correspond to the actual
probability function  only if the jet response were gaussian. 
  A detailed experimental simulation will be needed in order to
assess the validity of the obtained results in the real experimental
situation.\par  We now show $\log {\cal L}$ in the
$(\mgl-\msb,\mgl)$ plane for a few events where the $bb\ell\ell$
events originates from the cascade decay of Eq.~(\ref{bbll}) at
SPS1a. We calculate $\cal{L}$ using the following
procedure.  For each event in our
sample, characterized by a ($p_1,p_2$) pair of measured 
momenta for the $b$-jets, %we  define a Monte Carlo experiment.
we generate Monte Carlo events where the two $b$-jets have momenta
($p'_1$, $p'_2$),                      
where $p'_1$ and $p'_2$ are
randomly generated according to the function $\epsilon(p_1:p'_1)
\times\epsilon(p_2:p'_2)$. 
Each generated event corresponds to a curve in the 
($m_{\tilde{g}}$, $m_{\tilde{b}}$) plane which satisfies the 
equation $f(m_{\tilde{g}}, m_{\tilde{b}}, p_1,p_2)=0$. 
We histogram of the number of curves that go through each bin of a
$1\times1$~GeV grid in the $(\mgl,\msb)$ plane, 
for $n$ Monte Carlo events, normalized by dividing 
the bin contents by $n$.
%We define a $1\times1$~GeV grid in the $(\mgl,\msb)$
%plane, and for each experiment we add one hit to the 
% $(\mgl+\Delta\mgl,\msb+\Delta\msb)$ bins inside 
%which $F(\mgl,\msb)$ takes the value 1. 
%We fill the histogram  by generating $n$ Monte Carlo experiments, and we 
%normalize the function by dividing the bin content by $n$.
In the limit $n\rightarrow \infty$, this corresponds to 
\begin{equation}
{\cal L}(\mgl,\mgl+\Delta\mgl; \msb,\msb+\Delta\msb)
=\int Dp'_1 \int Dp'_2 \epsilon(p_1:p'_1)\epsilon(p_2:p'_2)
\theta(p'_1,p'_2,\mgl,\mgl+\Delta\mgl; \msb,\msb+\Delta\msb)
\label{likeii}
\end{equation}
where $\theta(p'_1,p'_2,\mgl,\mgl+\Delta\mgl; \msb,\msb+\Delta\msb)$
is 1 when the solution of Eq.~(\ref{glsb}) for the two 
$b$-jet momenta $p_1'$ and $p_2'$ goes through 
$(\mgl,\mgl+\Delta\mgl; \msb,\msb+\Delta\msb)$ and otherwise 
0. We take $n=10000$ for our calculations.

\begin{figure}
\includegraphics[width=5cm]{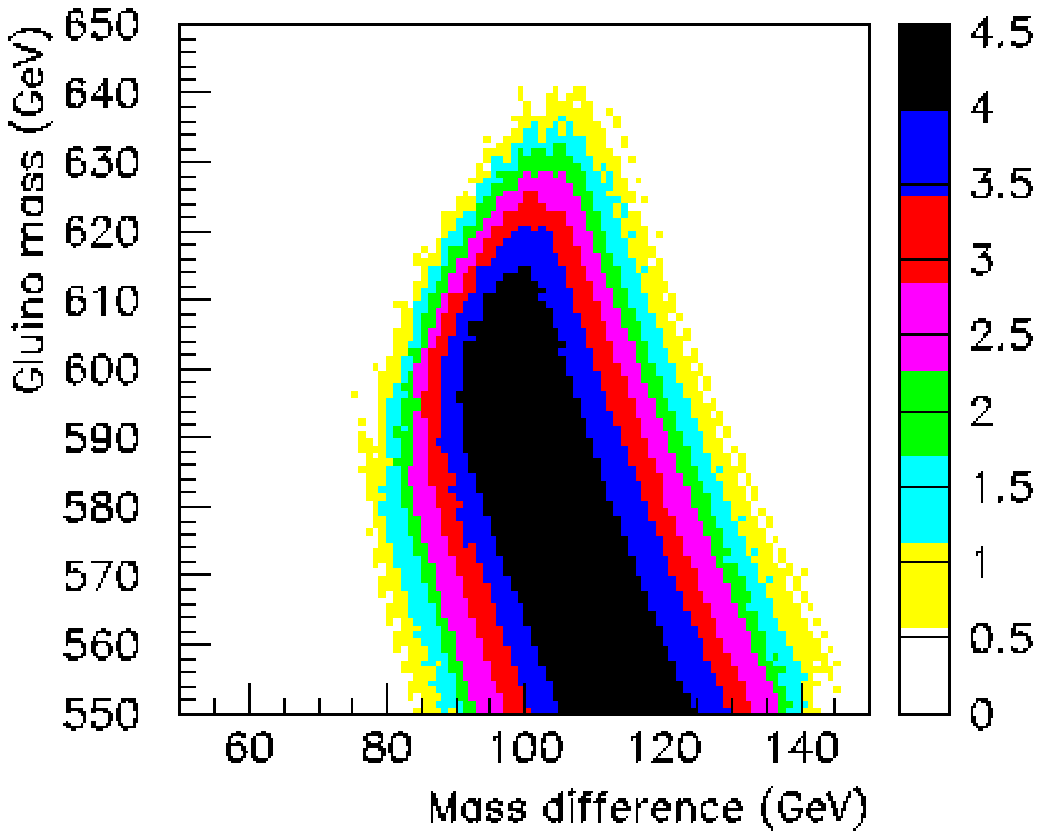}
\includegraphics[width=5cm]{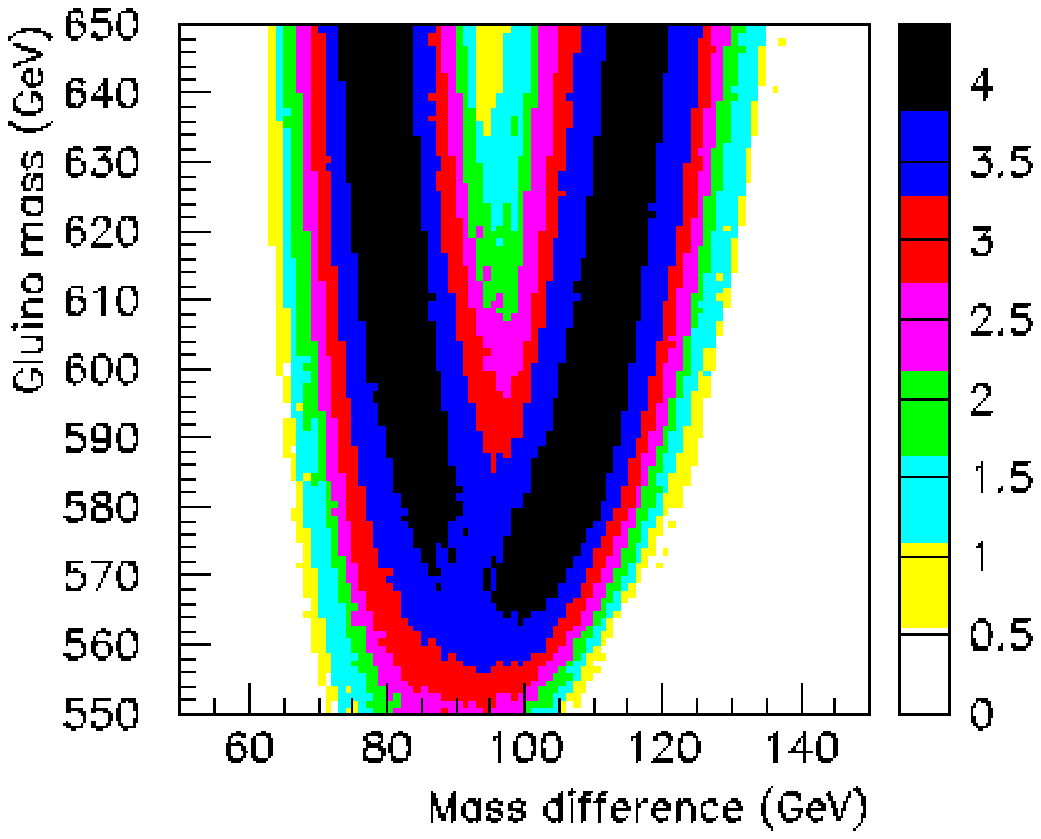}
\includegraphics[width=5cm]{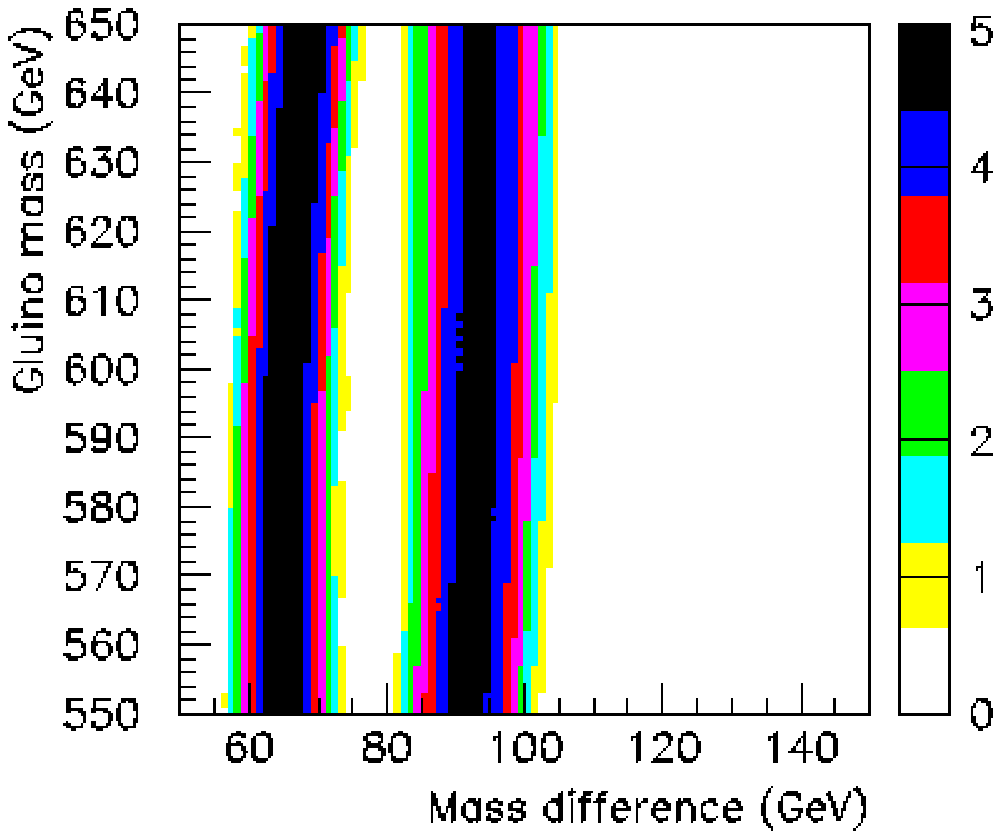}
\\
\begin{center}
a)\hskip 5cm b) \hskip 5cm c) 
\end{center}
\includegraphics[width=5cm]{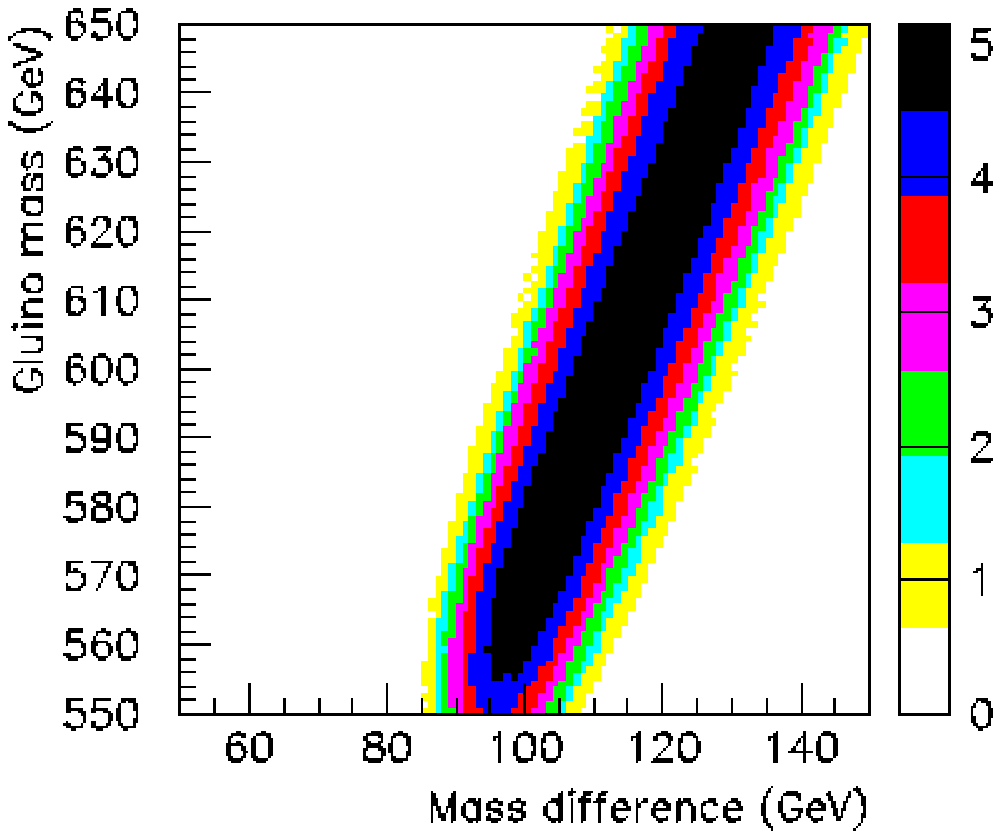}
\includegraphics[width=5cm]{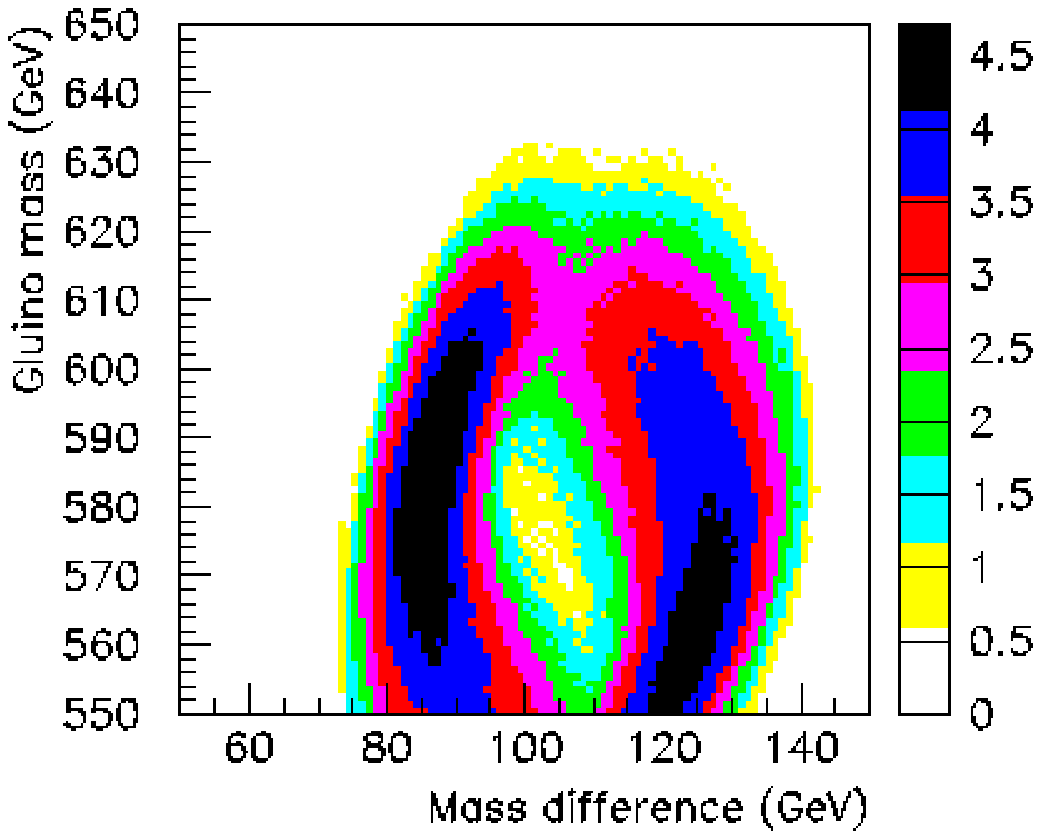}
\includegraphics[width=5cm]{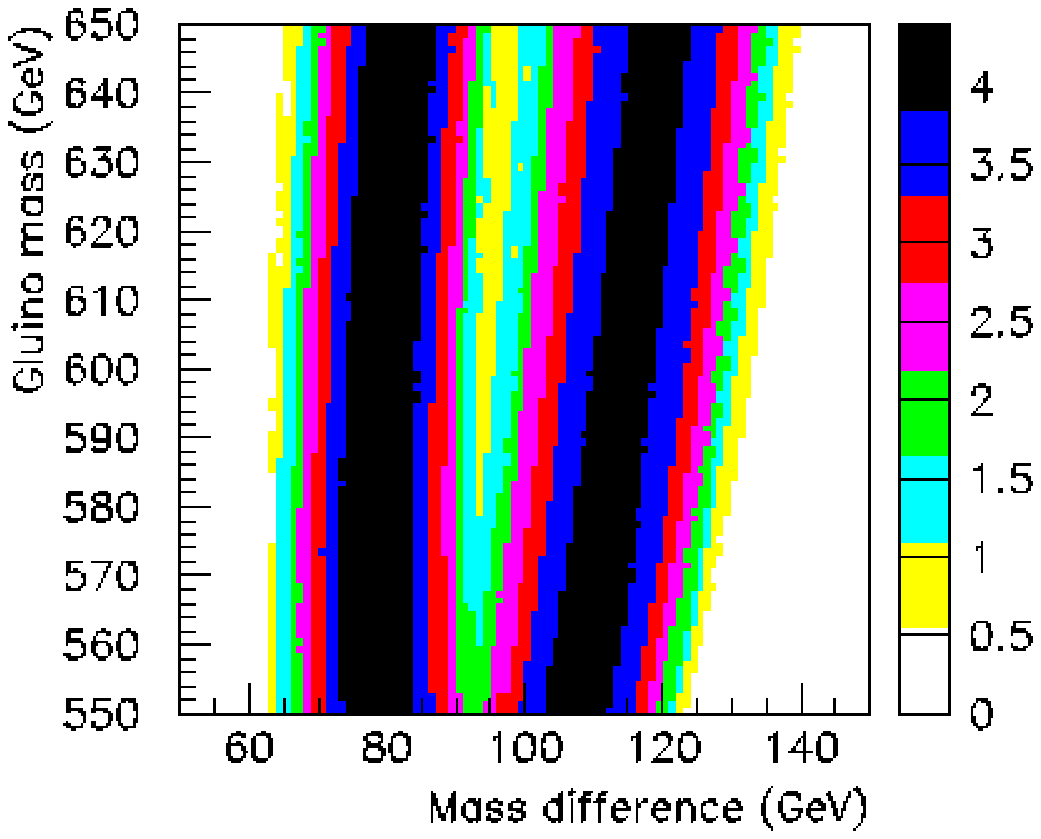}
\\
\begin{center}
d)\hskip 5cm e) \hskip 5cm f) 
\end{center}
\caption{Likelihood distributions in the $(\mgl-\msb,\mgl)$
for selected events.}
\label{events}
\end{figure}

In Fig. \ref{events},
we plot
\begin{equation}
\Delta \log {\cal L}
= \log( {\cal L}( \mgl, \mgl+\Delta\mgl, \msb,\msb+\Delta\msb)+ c)-
\log({\cal L}({\rm min})).
\label{evlik}
\end{equation}
where $c=0.001$ is a constant cutoff factor, which is 
needed as for each event we generate only 
a finite number of Monte Carlo experiments, 
and therefore some bins can have zero hits. 
The shape of the probability density distribution 
is different event by event, as it depends on the event kinematics.
For a few  events the  density distribution is 
parallel to the $y$ axis, therefore it only has  sensitivity to the 
$\mgl-\msb$ difference, but little sensitivity to the absolute value of 
the gluino mass.
The size of the band with significant probability is also different 
event by event, which means that some events will have 
more weight in the determination of the mass parameters. 

For each event, one defines in this way 
curves of equal probability in the ($\mgl,\msb$) plane. 
By combining the probabilities for different events,
a region of maximum probability in the ($\mgl,\msb$) is found,
where the curves of maximum probability for all events 
approximately cross. Given the fact that the curves have different
shapes for different events, the region thus defined has a limited
size, for an adequate number of events.
This region can be taken as a measurement for $\mgl$ and $\msb$.
We perform the combination as the 
product of ${\cal L}$ for the all events, and 
we define
\begin{equation}
\log {\cal L}_{\rm comb}(\mgl,\msb)\equiv  \sum_{\rm events} \log {\cal L}(\mgl,\msb)
\label{likelihood}
\end{equation}
As an estimator of the probability for a given ($\mgl,\msb$) pair, 
one can use
\begin{equation}
\Delta\chi^2(\mgl,\msb)\sim \Delta \log {\cal L}\equiv
\log {\cal L}_{\rm max}-\log {\cal L}_{\rm comb}(\mgl,\msb)
\end{equation}
Given the approximations introduced this does not however 
correspond to the statistical definition of $\Delta\chi^2$.

\subsection{Event analysis}
By following the procedure described in the previous subsection
we can build the combined likelihood for all the events 
defined as: 
\begin{equation}
\log {\cal L}_{\rm comb} ( \mgl, \mgl+\Delta\mgl,
\msb,\msb+\Delta\msb) = \sum_{\rm events} 
\log( {\cal L}( \mgl, \mgl+\Delta\mgl, \msb,\msb+\Delta\msb)+ c).
\end{equation}
As in Eq.~(\ref{evlik}) we have introduced a constant cutoff parameter
$c=0.001$ and  the  number of Monte Carlo
experiments used to build the likelihood for each event $n=10000$.

As we can see in Fig.~\ref{mglplot}, there are significant backgrounds
from accidental leptons.  The background subtraction must be carried
out using events with OSOF lepton pairs.  The correct log likelihood
function is schematically expressed as
\begin{equation}
\log {\cal L}_{\rm total}( {\bf m}_{\tilde{g}}, {\bf m}_{\tilde{b}},.....)
= \sum_{\rm OSSF} \log {\cal L}( {\bf m}_{\rm sig}, {\bf m}_{\rm bg})
+\sum_{\rm OSOF}  {\cal L}( {\bf m}_{\rm bg})
\end{equation}
where ${\bf m}_{\rm sig}$ express the parameters relevant to the
signal distribution, such as the masses of the sparticles involved in
the cascade decay, the decay branching ratios and so on. On the other
hand ${\bf m}_{\rm bg}$ are all the other parameters relevant to the
OSSF and OSOF events. This is rather complex procedure which is out of
the scope of this paper. Instead, we take the difference of the
functions for OSSF and OSOF lepton pair events
\begin{equation}
\log {\cal L}_{\rm sub}
\equiv \log {\cal L}_{\rm OSSF} - \log {\cal L}_{\rm OSOF}
\equiv
\sum_{\rm OSSF}\log{\cal L} 
-\sum_{\rm OSOF}\log{\cal L}.
\label{sub}
\end{equation}
In the limit of infinite statistics, $\log {\cal L}_{\rm sub}$ should
be independent from the contribution of accidental lepton pairs.
Therefore we use $\log{\cal L}_{\rm sub}$ in this paper.

\begin{figure}
\includegraphics[width=12cm]{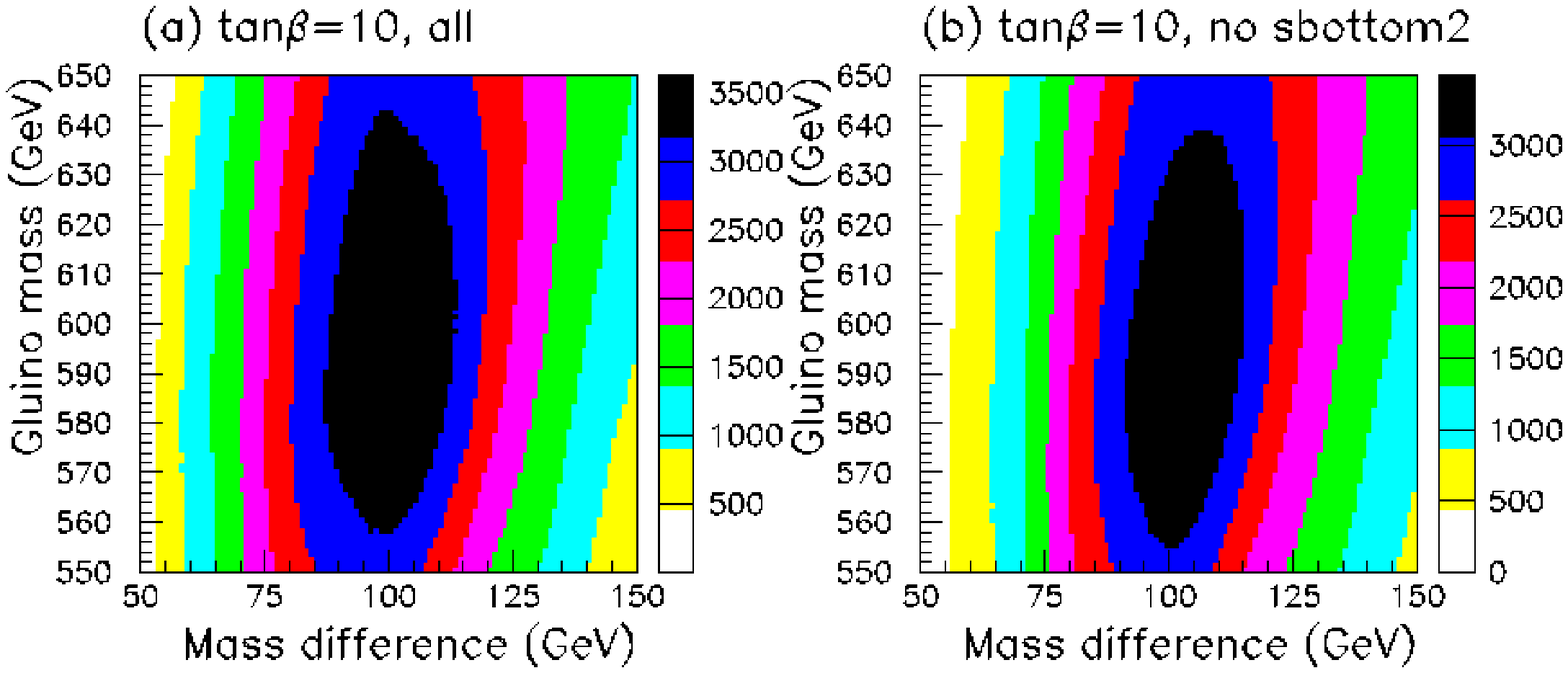}
\includegraphics[width=12cm]{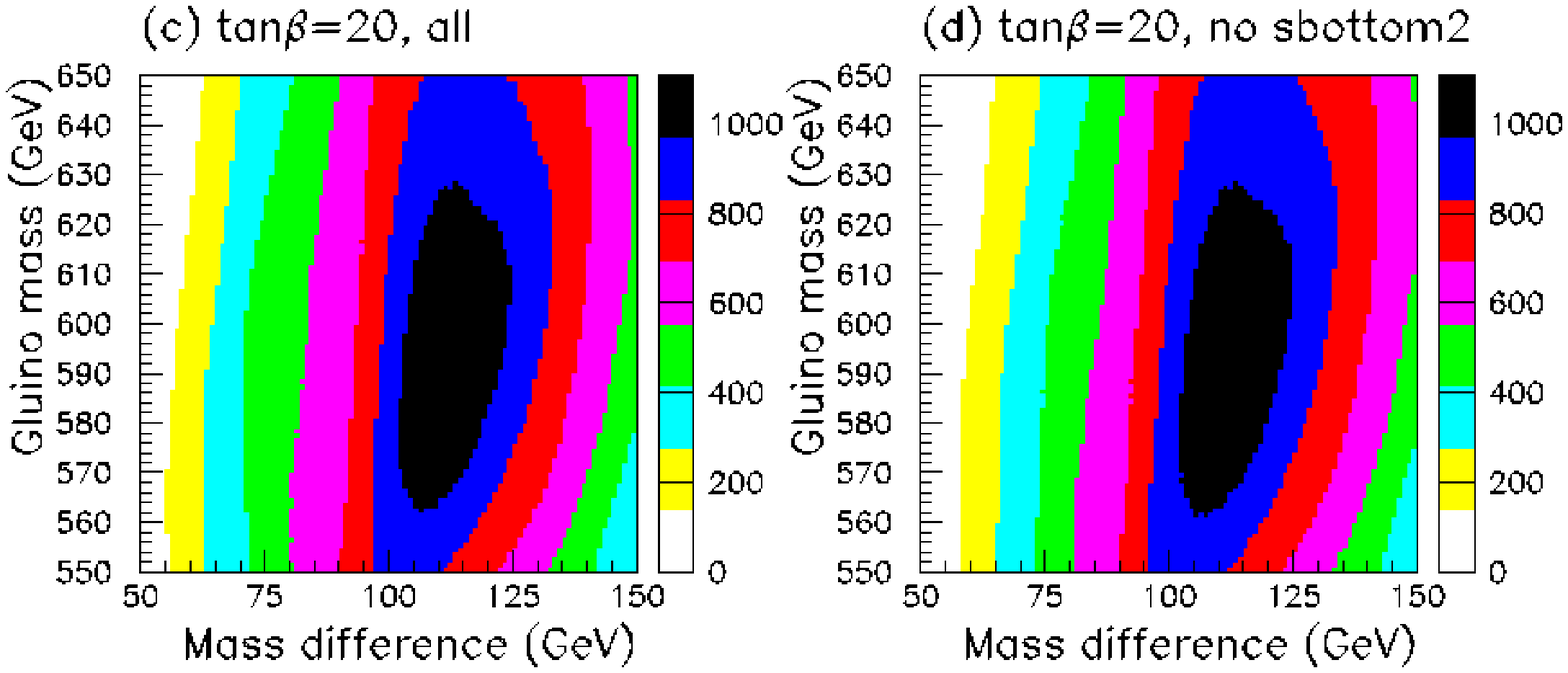}
\caption{Contours of the likelihood function
$\log {\cal L}_{\rm sub}$ in the $(\mgl-\msb,\mgl)$ plane:
(a) and (b) for $\tan\beta=10$ and (c) and (d) for $\tan\beta=20$,
respectively. 
The contours (b) and (d) are made without $\tilde{b}_2$ contributions.}
\label{nocut}
\end{figure}

We plot the contours of the function $\log {\cal L}_{\rm sub}$ in
Fig.~\ref{nocut}, where plots (a) and (b) [(c) and (d)] are for
$\tan\beta=10$ [$\tan\beta=20$].  The distributions (a) and (c) are
produced accepting all the events which pass the selections, whereas
distributions (b) and (d) are produced using an event sample where the
events including a $\tilde{b}_2$ decay have been rejected.

In Fig.~\ref{nocut} (a) and (c), the position of the peak for
$\mgl-\msb$ is roughly consistent with the input value.  Unlike the
gluino and sbottom mass fits in the previous section, we obtain the
correct peak position without the need of artificially choosing among
multiple solutions.  The likelihood distribution can be used to
determine the $\tilde{g}$ and $\tilde{b}$.  We restrict the likelihood
distribution for $591$~GeV $<m_{\tilde{g}}<599$~GeV(within 4~GeV from
the input gluino mass). We then fit the distribution around the peak
assuming gaussian distribution, The likelihood distribution peaks at
the gluino and sbottom mass difference as 99.5~GeV for $\tan\beta=10$,
104.2~GeV for $\tan\beta=15$, and 113.9~GeV for $\tan\beta=20$, where
the input value is 103.3~GeV, 109.9~GeV and 116.5~GeV, respectively.
The fitted values display shift of about 4~GeV from the true value. We
ascribe this effect to our simplified modeling of the jet smearing in
building the likelihood function, which should disappear once the
detector response is properly taken into account in the unfolding
procedure.

\begin{figure}
\includegraphics[width=12cm]{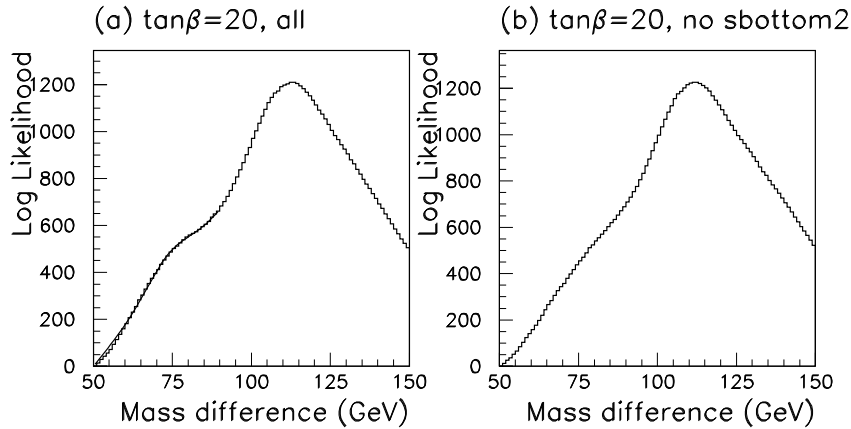}
\caption{The likelihood as a function of $\mgl-\msb$
for $\tan\beta=20$:
(a) for all events and
(b) without $\tilde{b}_2$ events.}
\label{tanb20}
\end{figure}

By comparing the left side with the right side of Fig~\ref{nocut}, we
also observe a slight shift in the position of the maxima of the
distributions, showing that the distributions are sensitive to the
presence of $\tilde{b}_2$ decays.  The $\tilde{b}_2$ contribution
however manifests itself in Fig.~\ref{nocut} (a) and (c) only as a
flattening of the distribution around $\mgl-\msb = 70$~GeV at
$m_{\tilde{g}}\sim 595$~GeV.  No secondary peak can be observed
because of the experimental smearing, and of the fact that the
branching ratio into $\tilde{b}_2$ is much smaller.
In Fig.~\ref{tanb20}(a), we show the distribution of $\log {\cal
L}_{\rm sub}$ as a function of $\mgl-\msb$ at $\tan\beta=20$,
restricting the gluino mass in the region $591$~GeV$< \mgl<$599~GeV
again.  On the left of the peak corresponding to the $\tilde{b}_1$
mass, we see a small bump in the distribution.  This bump is not
observed in the mass distribution made without $\tilde{b}_2$
contribution (Fig.~\ref{tanb20}(b)).  In order to claim the presence
of a second component in the distribution on the data, the ability of
correctly reproducing the likelihood distribution for $\tilde{b}_1$
events would be needed.  It is also difficult to extract a statistical
significance for the $\tilde{b}_2$ shoulder as our definition of the
likelihood function is approximate one, and we did not treated the
background subtraction correctly as can be seen in Eq.~(\ref{sub}).

In Fig.~\ref{smallmsb}, it is rather hard to see the effect of
$\tilde{b}_2$ unlike in Fig.~\ref{tanb20}.  The apparent discrepancy
probably comes from the fact that the likelihood analysis is more
sensitive to the model parameter than simply solving Eq.~(\ref{glsb})
for a fixed gluino mass.  The likelihood analysis not only takes care
of the most plausible value of the sbottom mass for a fixed gluino
mass, but also includes possible statistical fluctuations, which vary
event by event as seen in Fig.~\ref{events}.  For example, badly
mis-measured events have less chance to be consistent with the input
gluino mass, providing a natural cut for the event selection.

\begin{figure}
\includegraphics[width=12cm]{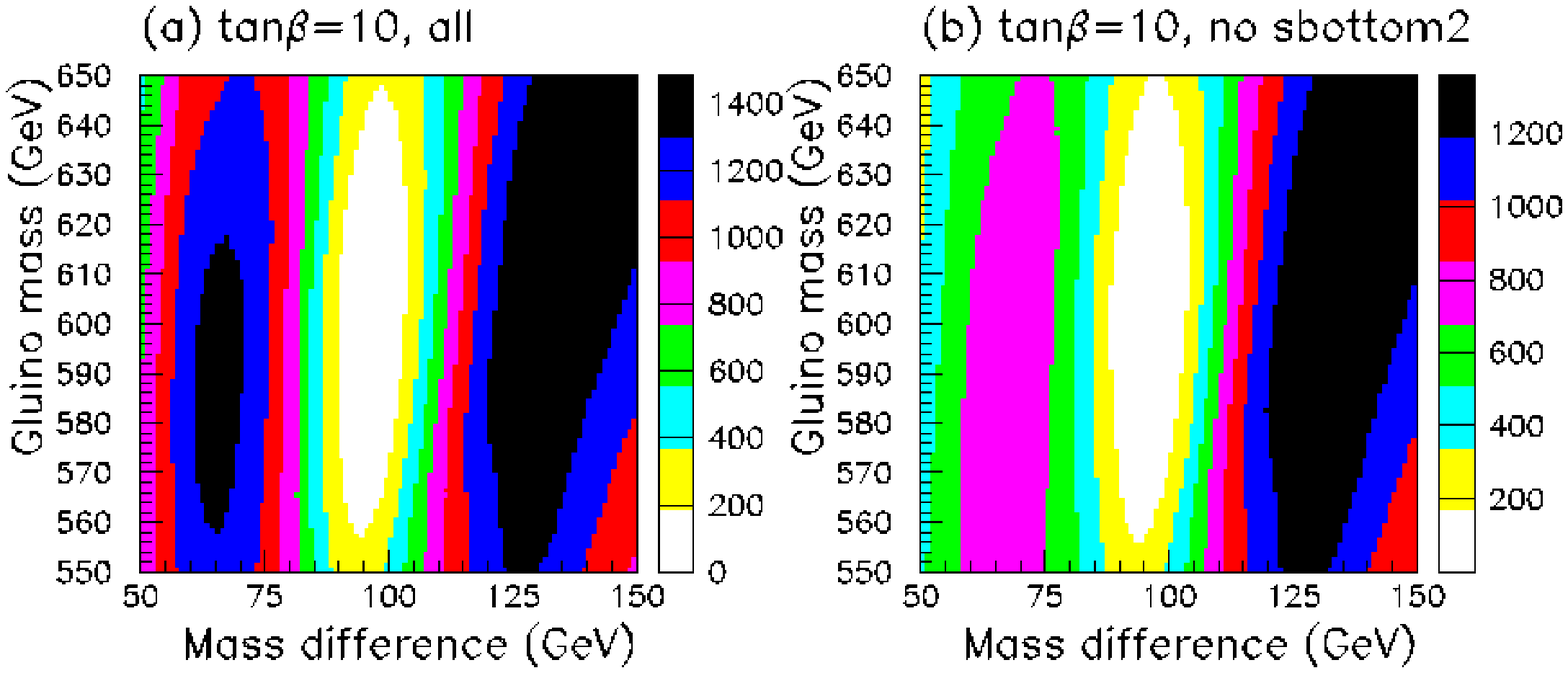}
\includegraphics[width=12cm]{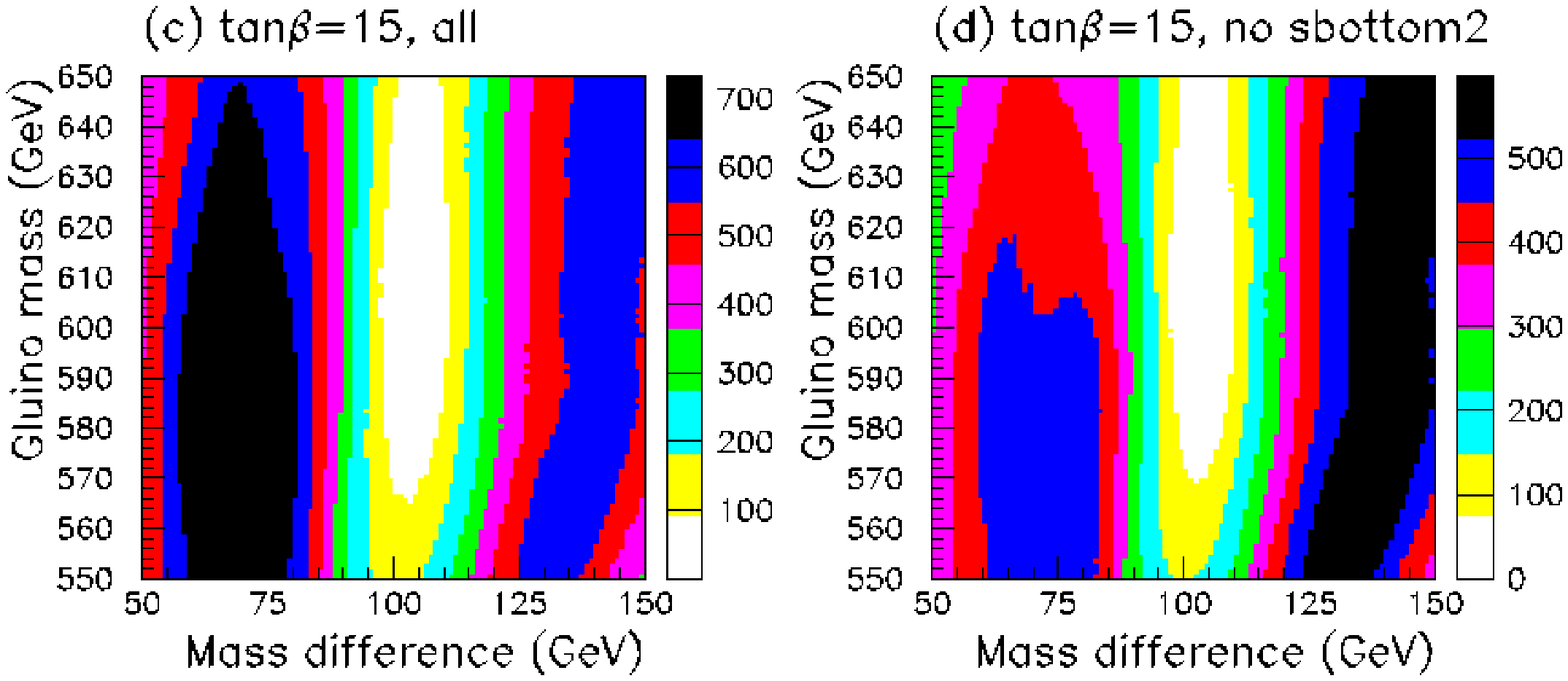}
\caption{The contours of the likelihood function
in the $(\mgl-\msb$, $\mgl)$ plane after 
the cut given in Eq.~(\ref{cut}):
(a) and (b) are with and without $\tilde{b}_2$ events
at $\tan\beta=10$,
respectively,
and (c) and (d) at $\tan\beta=15$.}
\label{aftercut2}
\end{figure}

\begin{figure}
\includegraphics[width=6cm]{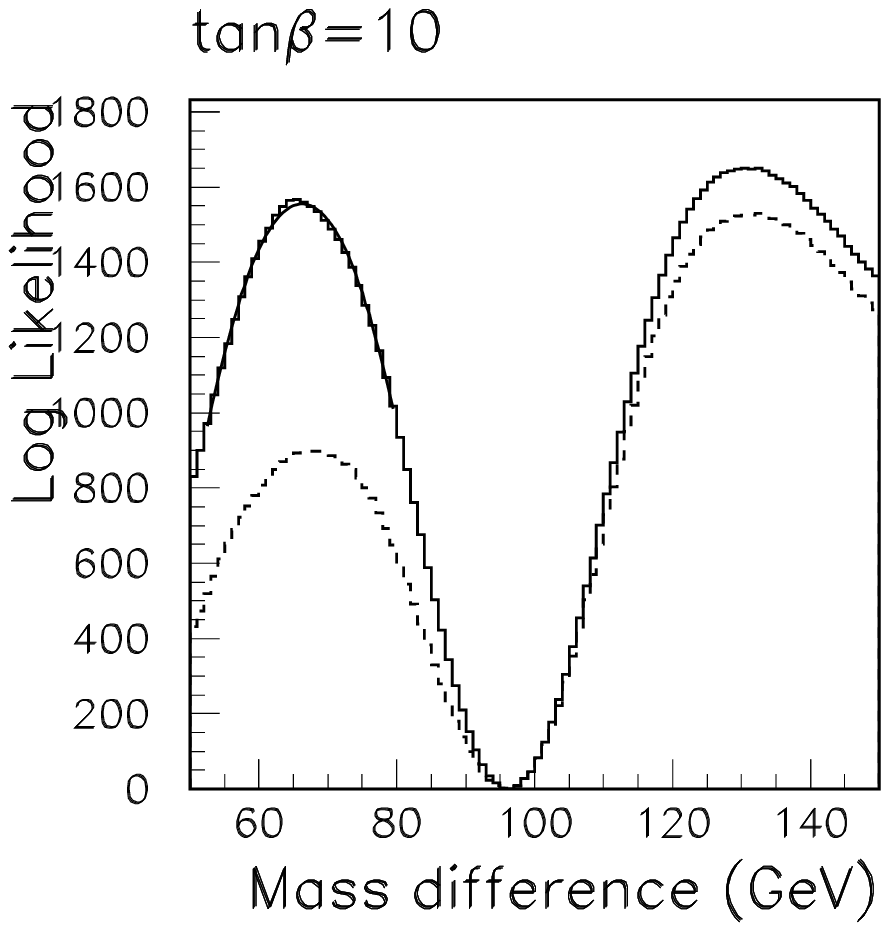}
\includegraphics[width=6cm]{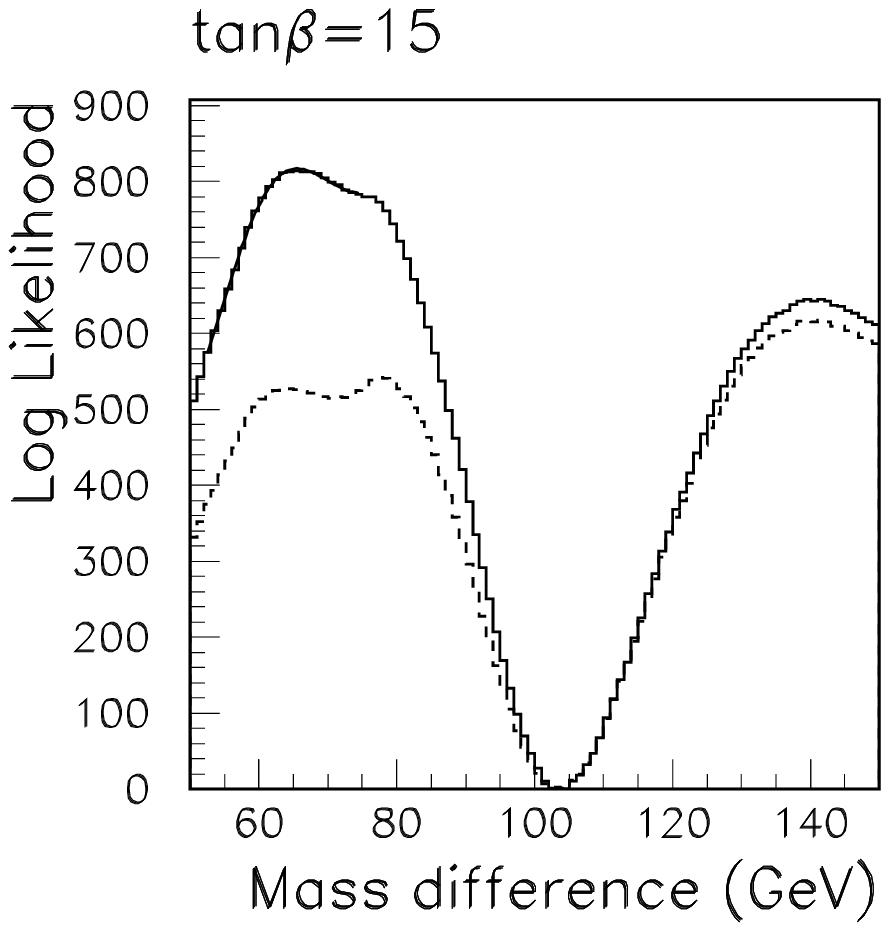}
\begin{center}
%a)\hskip 5cm b)
\end{center}
\caption{The likelihood as a function of $\mgl-\msb$
with the 
cut given in Eq.~(\ref{cut}):
(left) for $\tan\beta=10$
and (right) for $\tan\beta=15$.}
\label{aftercut1}
\end{figure}

A significant part of the background under the $\tilde{b}_2$ is the
tail of the smeared probability distribution for events which
correctly reconstruct the $\tilde{b}_1$ mass.  One can therefore try
to remove from the distributions the events consistent with
$\tilde{b}_1$ in order to improve the signal to background ratio for
$\tilde{b}_2$.  To reduce the $\tilde{b}_1$ events, an event is
required to satisfy the condition
\begin{equation}
\sum_{\rm cut\ region} {\cal L}< {\cal L}_{\rm cut}\,
\label{cut}
\end{equation}
where the sum is made for bins in a cut region in the 
$({\mgl-\msb,\mgl})$ plane.
We choose the region  as
\mbox{$550$~GeV$<m_{\tilde{g}}<$ $650$~GeV},
and $m({\rm min})$ $<m_{\tilde{g}}-m_{\tilde{b}}<$ $m({\rm max})$,
which corresponds to the region around the $\tilde{b}_1$ peak.  We use
the cut value ${\cal L}_{\rm cut}=20$.  The relevant $m({\rm min})$
and $m({\rm max})$ values are listed in Table~\ref{afterfit}.  The
contours of $\log {\cal L}_{\rm sub}$ after this cut are shown in
Fig.~\ref{aftercut2}: (a) and (b) for $\tan\beta=10$ and (c) and (d)
for $\tan\beta=15$.  The contours (b) and (d) are made without the
$\tilde{b}_2$ contribution.  By comparing the contours with and
without the $\tilde{b}_{2}$ contribution, the presence of the
$\tilde{b}_{2}$ can be clearly observed in the plots.  
In Fig.~\ref{aftercut1}, we again plot the likelihood function as the function
of $\mgl-\msb$, for \mbox{$550$~GeV$<m_{\tilde{g}}<$ $650$~GeV}.  The
distributions without the $\tilde{b}_2$ contribution are also shown as
dashed histograms.  The signal to background ratio is much improved
with respect to what is seen on the left side of Fig.~\ref{tanb20}, and
it is about 1:1. However, from an inspection of the mass distribution
event by event, the purity of the signal after the likelihood cut 
does not appear significantly improved with respect to Fig.~\ref{smallmsb}.
Moreover the position of the peak corresponding to
$\tilde{b}_{2}$ is dependent from the cuts applied. This is
illustrated in Table~\ref{afterfit}, where the results of fits to the
peak position are shown for two different values of the applied cuts
both for the full sample and for the pure $\tilde{b}_{2}$ signal.
Even when only considering the $\tilde{b}_{2}$ signal, the peak
position depends on the cuts, albeit with a milder dependence than for
the full sample.  It will therefore be problematic to extract a mass
measurement from Fig.~\ref{aftercut1}, even assuming a priori the
existence of a $\tilde{b}_{2}$ contribution.
 
\begin{table}
\begin{tabular}{|c|c|c|c|c|}
\hline
&$m$(min)&$m$(max)&	all	&	$\tilde{b}_2$ only \cr
\hline
$\tan\beta=10$ &87&108&	64.1 $\pm$ 0.2&	62.7 $\pm$ 0.3\cr
                      &82&113&	60.3 $\pm$ 0.3&	60.2 $\pm$ 0.5\cr
$\tan\beta=15$ &94&115&	61.1 $\pm$ 0.7&	65.3 $\pm$ 0.4\cr
                      &89&120&	62.7 $\pm$ 0.3&	62.6 $\pm$ 0.6\cr
\hline
\end{tabular}
\caption{The peak positions at the smaller $\mgl-\msb$ region in
Fig.~\ref{aftercut1} for $\tan\beta=10$ and 15. The peak positions
for the signal $\tilde{b}_2$ distribution are also shown.}
\label{afterfit}
\end{table}

\section{Discussion}
\label{sec6}

Supersymmetric models predict the existence of heavy 
superpartners which decay subsequently into 
the lighter superpartners. The lightest SUSY particle 
is stable and neutral, and escapes detection. 
Therefore two undetected particles will be present in
each event. Moreover the partonic center of mass energy
is unknown for hadron collisions. As a result, the
complete kinematic reconstruction of SUSY events at hadron
colliders is problematic.

We propose a new analysis method to solve the decay kinematics of this
decay at the LHC, based on imposing the on-shell condition on the
momenta of the particles participating in the cascade.  The single
cascade decay is solvable if a decay chain consisting of at least 4
successive two-body decays, involving 5 sparticles can be identified.
In this case each event defines the 4 dim hypersurface in the 5 dim
sparticle mass space.  The potential of this method should be compared
to the method previously used for this analysis based on the
measurement of kinematic edges of invariant mass combinations of the
detected decay particles.

The merits and demerits of the new method may be summarized as follows:
\begin{itemize}
\item The cascade decay is solved based on the exact 
formulae.
\item Mass peaks would be reconstructed, as opposed
to the kinematic edges. This allows us to perform measurements 
even if the significant backgrounds exist or statistics 
are small. Note that 
backgrounds will  not exhibit  peaks corresponding to the
signal region.  
\item In the case where sfermion masses are heavier than 
gaugino masses, 
the cascade decays are expected to be shorter, and one 
can not therefore use this method. 
The endpoint method provides a mass information even 
in this case. 
\end{itemize}

We note that in case two decay chains can be simultaneously identified
in the event, the method can be applied to shorter decay chains,
consisting of only three decays each. In fact in this case two
additional constraints can be applied by requiring that the sums of
the transverse components of the momenta of the two lightest
neutralinos equal the two components of the measured missing
transverse momentum.

Further on, if one sparticle cascade decay is solved by the mass
relation method, the candidate LSP momentum $p_T$ would be obtained.
We can then calculate the transverse momentum of the other LSP $p'_T$
as:
\begin{equation}
p'_T= -p_T+ P_{\rm miss}. 
\end{equation}
For the cascade decay to which the second LSP belongs only two 
components of the neutralino four momentum are unknown, 
therefore a cascade decay with $n_{decay}\ge 3$ 
can be solved. 

To see the performance of the mass relation method, we have studied in
this paper the problem of measuring sparticle masses in the cascade
decay: $
\tilde{g}\rightarrow 
\tilde{b}b_2 \rightarrow 
\tilde{\chi}^0_2 b_1b_2 \rightarrow
\tilde{\ell}b_1b_2\ell_2 \rightarrow 
\tilde{\chi}^0_1b_1b_2\ell_1\ell_2$.
We have performed a detailed study for some benchmark SUSY model
points including backgrounds and a parameterized simulation of
detector effects.  We performed the exercise in a simplified fashion,
by fixing the masses of the three lighter particles to avoid the
practical complications in handling a large number of parameters.
Then each event becomes an allowed curve in the gluino-sbottom mass
plane, passing through the point corresponding to the true gluino and
sbottom masses.  We first addressed the mass reconstruction through
the ``event pair analysis'', which determines the sparticle masses
from the distribution of the solutions of any event pairs in the
selected sample.  The method reconstructs $\msbi$ correctly for SPS1a
where $\tan\beta$ is varied from 10 to 20.  Note that the signal
branching ratio becomes a factor 4 smaller for $\tan\beta=20$ with
respect to $\tan\beta=10$, but the $\tilde{b}_1$ peak is still clearly
observable.  On the other hand, in order to obtain the correct mass,
one needs to artificially choose among the multiple available
solutions.

A more global approach requires the usage of all available events
simultaneously. For this approach we constructed an approximate
likelihood function for the true gluino and sbottom masses, taking
into account the experimental smearing in the measurement of the
$b$-jets.  When the peak position of the likelihood distribution is
used to extract the mass, the $\tilde{b}_1$ mass is measured without
the problem of the multiple solutions of the event pair analysis.

We also try to probe the presence of a $\tilde{b}_2$ in our Monte
Carlo sample.  The sbottom mass matrix is parameterized by three
parameters $m_{\tilde{b}_1}$, $m_{\tilde{b}_2}$ and the mixing angle
$\theta$.  Successful extraction of the $\tilde{b}_2$ would be an
important step to fully understand the nature of the third generation
sparticles.  No clear result is achieved for $\tilde{b}_2$, as the
branching ratio into $\tilde{b}_2$ is small for the parameters we have
chosen, and also the small difference between the two sbottom states
is comparable to the resolution in the experimental measurement of jet
momenta.  It is however clear that, even with the small statistics
available for the case $\tan\beta=20$, a hint for the deviation from a
single-mass case can be seen in the distribution.

The physics output of our analysis is therefore the possibility to
extract information on the third generation sector, even for rather
small input statistics.  The measurement of the third generation
sparticle masses are important theoretically. In mSUGRA,
$\tilde{b}_{L(R)}$ masses are same as the other sparticle masses at
the GUT scale but non-universality is induced by RGE running at the
weak scale due to the Yukawa coupling. In addition to that, the
left-right mixing of sbottom is induced by the $F$ term of the
Superpotential which is proportional to $\mu\tan\beta$. The mass shift
around 20~GeV from $\tan\beta=10$ and $\tan\beta=20$ is due to the
mixing effect.  Our method is sensitive to the $\tan\beta$ dependence
as it could yield a sensitivity to the mass difference between the two
states of the order of a few GeV.

\section*{Acknowledgments}
We thank members of the ATLAS Collaboration for helpful
discussions. We have made use of ATLAS physics analysis and simulation
tools which are the result of collaboration-wide efforts. 
This work is
supported in part by the Grant-in-Aid for Science Research, Ministry
of Education, Science and Culture, Japan
(No.~11207101 and 15340076 for K.K. and
No.~14540260,14046210 and  16081207 for M.M.N.). 
M.M.N. is also supported in part
by a Grant-in-Aid for the 21st Century COE ``Center for
Diversity and Universality in Physics''.

\section*{Appendix} 
 
As described in section 2, we have solved the cascade decay 
in Eq.~(\ref{bbll})  by expanding the the lightest neutralino momentum 
by the momenta of $l_1$, $l_2$ and $b_1$ as in Eq.~(\ref{expand})
\begin{eqnarray}
&&\vnui=a \vli+ b \vlii+ c\vbi
\nonumber
\end{eqnarray}
The parameters $(a,b,c)$ may be written as the function of 
sparticle masses and the LSP energy as:
\begin{equation}
\left(
\begin{array}{c}
a\cr
b\cr
c\cr
\end{array}
\right)
= {\cal M}^{-1}\left[[{\xno}]\mnui^2 +[{\xni}]\msb^2+[{\xnii}] \mnui \enui\right] 
={\cal M}^{-1}{\cal X} \left(
\begin{array}{c}
\mnui^2\cr
\msb^2\cr
\mnui \enui
\end{array}
\right),
\end{equation}
where 
\begin{eqnarray}
{\cal X}&=&[\xno,\xni,\xnii],\cr
[\xno]&=&\frac{1}{2m^2_{\tilde{\chi}^0_1}}\left(
\begin{array}{l}
-\msl^2+\mnui^2\cr
-\mnuii^2+\msl^2+2\pli\cdot\plii\cr
\mnuii^2+m_b^2+2\pbi\cdot (\pli+\plii)
\end{array}
\right),
\cr
[\xni]&=& \left(
\begin{array}{c}
0\cr 0\cr -1/2 \end{array}\right),
\ \ [\xnii]= 
\frac{1}{\mnui}\left(
\begin{array}{c}\eli\cr
\elii\cr\ebi\end{array}\right).
\end{eqnarray}
and $M$ is defined already in section 2 as:
\begin{eqnarray}
{\cal M}= \left(
\begin{array}{ccc}
\vli\cdot\vli &\vli\cdot\vlii& \vli\cdot\vbi\cr
\vli\cdot\vlii & \vlii\cdot\vlii & \vlii \cdot\vbi\cr
\vli \cdot \vbi & \vlii\cdot \vbi & \vbi\cdot\vbi\cr 
\end{array}
\right).
\nonumber
\end{eqnarray}

By using the on-shell condition of the neutralino mass 
\begin{equation}
\enui^2=(a,\ b,\ c){\cal M} \left(\begin{array}{c}a\cr b\cr c
\end{array}\right) +\mnui^2,
\end{equation}
we obtain the following equation:
\begin{eqnarray}
&&A_{33}\left(\frac{\enui}{\mnui}\right)^2+ 
2\left(A_{13} + \frac{\msb^2}{\mnui^2} A_{23}\right)
\left(\frac{\enui}{\mnui}\right)
+ \left(A_{11}+ 2 \frac{\msb^2}{\mnui^2} A_{12}+ 
\frac{\msb^4}{\mnui^4}A_{22}\right)
=0,   
\nonumber
\end{eqnarray}
where $A_{ij} = [{\rm x_i}]^T {\cal M}^{-1} [{\rm x_j}]-\delta_{ij}(i-2)/\mnui^2$
The solution of $\enui$ is 
expressed as 
\begin{eqnarray}
\enui&=&\frac{\mnui}{A_{33}}\left(-A_{13} - \frac{\msb^2}{\mnui^2} A_{23}
\pm \sqrt{D}\right),
\cr
\cr
D&=&\left(A_{13} + \frac{\msb^2}{\mnui^2} A_{23}\right)^2
-A_{33}\left(A_{11}+ 2 \frac{\msb^2}{\mnui^2} A_{12}+ 
\frac{\msb^4}{\mnui^4}A_{22}\right)
\cr
&=&\left(\frac{\msb^2}{\mnui^2}\right)^2\left(A_{23}^2-A_{33}A_{22}\right)
+ 2\left(\frac{\msb^2}{\mnui^2}\right)\left(A_{13}A_{23}-A_{33}A_{12}\right)
\cr
\cr
\cr
&&+\left(A_{13}^2-A_{11}A_{33}\right).
\end{eqnarray}

%It is mathematically trivial to follow the above 
%steps. However, in an actual experiment, momentum 
%measurement involves numerical errors, which 
%may not be negligible to for the purpose to reconstruct 
%the small LSP momentum. In MSUGRA model, the mass difference 
%$m_{\tilde{b_1}}-m_{\tilde{\chi}^0_2}$ is large, therefore 
%$\vert \vbi\vert\gg \vert\vli\vert$ or $\vert \vlii\vert$. 
%The errors for the reconstructed  LSP momentum  would be 
%dominated by the smearing of the $b$ jet momentum. 

When the sbottom comes from gluino decay, we can further 
use the  gluino mass shell condition: 
\begin{equation}
\mgl^2=2\pbii\cdot\pnui+2\pbii\cdot(\pbi+\plii+\pli)+\msb^2+m_b^2,
\end{equation}
where 
\begin{eqnarray}
\pbii\cdot\pnui&=&\enui\ebii-\vbii\cdot\vnui=\mnui^2 F_0+
F_1\msb^2+
F_2\mnui^2\left(\frac{\sqrt{D}}{A_{33}}\right),
\cr
F_0&\equiv&-\left(\frac{\ebii}{\mnui} -[{\cal K}_{b2}]^T\cdot[\xnii] \right)
\left(\frac{A_{13}}{A_{33}}\right)
-[{\cal K}_{b2}]^T[\xno],
\cr
F_1&\equiv&-\left(\frac{\ebii}{\mnui} -[{\cal K}_{b2}]^T\cdot[\xnii] \right)
\left(\frac{A_{23}}{A_{33}}\right)
-[{\cal K}_{b2}]^T[\xni],
\cr
F_2&\equiv&\pm\left(\frac{\ebii}{\mnui} -[{\cal K}_{b2}]^T\cdot[\xnii] \right),
\cr
[{\cal K}_{b2}]^T&\equiv&(\vbii\cdot\vli,\ \vbii\cdot\vlii,\ \vbii\cdot\vbi)
{\cal M}^{-1}. 
\end{eqnarray}
The equation involving the gluino and sbottom masses is of the form
\begin{eqnarray}
&&Q_{11} \mgl^4+2Q_{12} \mgl^2 \msb^2+ Q_{22}\msb^4
+2Q_{1} \mgl^2+ 2Q_{2}\msb^2 + Q=0,
\nonumber
\end{eqnarray}
where 
\begin{eqnarray}
Q_{11}&=&1,\ \  Q_{12}=-2F_1-1,
\cr
Q_{22}&=&(2F_1+1)^2-F_2^2\frac{A_{23}^2-A_{33}A_{22}}{A^2_{33}},
\cr
Q_{1}&=&-2F_0 \mnui^2-2\pbii\cdot(\pbi+\pli+\plii)-m_b^2,
\cr
Q_2&=&Q_{12}Q_{1}-F_2^2 \mnui^2\frac{A_{13}A_{23}-A_{33}A_{12}}{A^2_{33}},
\cr
Q&=&Q_{1}^2-F_2^2\mnui^4\frac{A_{13}^2-A_{11}A_{33}}{A^2_{33}}.
\end{eqnarray}
It should be noted that 
one can  derive a equation 
of the form $m_I^4 + C_1m_I^2+ C_2=0$ for any particle $I$ involved in
the cascade.

\end{document}